Accurate calculation of Stokes drag for point-particle tracking in two-way coupled flows

J.A.K. Horwitz[1] and A. Mani[1,*]

September 24, 2015

**Abstract**

In this work, we propose and test a method for calculating Stokes drag applicable to particle-laden fluid flows where two-way momentum coupling is important. In the point-particle formulation, particle dynamics are coupled to fluid dynamics via a source term that appears in the respective momentum equations. When the particle Reynolds number is small and the particle diameter is smaller than the fluid scales, it is common to approximate the momentum coupling source term as the Stokes drag. The Stokes drag force depends on the difference between the undisturbed fluid velocity evaluated at the particle location, and the particle velocity. However, owing to two-way coupling, the fluid velocity is modified in the neighborhood of a particle, relative to its undisturbed value. This causes the computed Stokes drag force to be underestimated in two-way coupled point-particle simulations. We develop estimates for the drag force error as function of the particle size relative to the grid size. We then develop a correction method that estimates the undisturbed fluid velocity from the computed disturbed velocity field. The correction scheme is tested for a particle settling in an otherwise quiescent fluid and is found to reduce the error in computed settling velocity by an order of magnitude compared with common interpolation schemes.

[1] Department of Mechanical Engineering, Stanford University
[*] Corresponding Author, alimani@stanford.edu

# 1. Introduction

Fluids laden with particles are abundant in environmental and industrial settings ranging from the atmospheric transport of volcanic ash, coal-fired powerplants, and soot formation in engines. It has been observed by many authors cf. [3, 29, 32] that fluid statistics in a turbulent flow are altered when particles are introduced. The no-slip condition at particle surfaces provides an additional dissipation mechanism. In particle boundary layers, the fluid velocity must change from the characteristic undisturbed value which is of order $u'_{rms}$ to the particle velocity. In fully resolved simulations of particle-turbulence interaction, Burton and Eaton [31] observed this variation may happen on the scale of a few particle diameters. Other primary questions in particle-laden flows concern the nature of particle dispersion and concentration where the distribution of particles may influence other physical processes. For example, in particle-based solar collectors, non-uniform particle concentration can impact the overall system efficiency [9]. Another example concerns fluidized bed reactors where particles act as catalyzing agents lowering the activation energy necessary for reactants to reach activated complex. However, inhomogeneities in the particle field owing to preferential concentration mean some portions of the flow will have reacted while other regions will have yet to undergo chemical reaction [11].

While there have been some efforts to study particle-laden flows using fully resolved simulations cf. [20, 30, 31], the number of grid points required to resolve particle boundary layers makes simulation of a large number of particles impractical. A popular simulation method is the Lagrangian point-particle method. In this method, particles are tracked in their respective Lagrangian frames while the fluid is simulated using standard Eulerian methods. Particles are represented as point sources of momentum and energy. Coupling between the particle and fluid phases is accomplished via Newton's third law. In the particle momentum equation, the magnitude of the force felt by the fluid owing to the particle is equal in magnitude and opposite in direction to the force experienced by the particle owing to the fluid.

In a fully resolved simulation, the force felt by the particle would be equal to the fluid stress integrated over the boundary of the particle. In a point-particle simulation, the fluid stresses are not computed at the scale of each particle so the resulting force must be modeled using nearby fluid information. For a particle suspended in a turbulent flow there is no general point force model that accounts for all of the fluid interactions that a particle experiences. In the limit of small particle Reynolds number, and particle diameter smaller than the Kolmogorov scale, the Maxey-Riley equation [18] provides a very general model for the interactions felt by a particle with a fluid. These effects include gravity, fluid acceleration, Stokes drag, Faxen corrections owing to a spatially non-uniform flow, added mass, and Basset-Boussinesque history force owing to unsteady effects. Calzavarini et al. [7] performed simulations of particle-laden flow and showed that point-particles obeying Stokes drag augmented by Faxen correction terms had no discernable change in the variance and kurtosis of their Lagrangian acceleration statistics compared with particles obeying Stokes drag alone for particle diameters less than four times the Kolmogorov scale. For gas-solid flows where the density ratio between the particle and fluid typically exceeds 1000, the leading order contributions in the particle momentum equation are the Stokes drag and gravity force.

For terrestrial applications, the gravitational force is safely modeled as a constant body force and requires no special treatment. The Stokes force however requires more consideration. Originally derived in 1850 [8], Stokes showed that the resistance felt by a particle moving slowly at constant velocity through an otherwise stationary viscous fluid was proportional to the particle's velocity. For a fluid that is non-stationary even in the absence of the particle, an equivalent statement is the drag force experienced by the particle is proportional to the difference between the particle velocity and the undisturbed fluid velocity prior to the introduction of the particle evaluated at the particle location. Mathematically, the Stokes' drag can be expressed as $F_d = m_p(\tilde{u} - v_p)/\tau_p$ where $m_p$ and $\tau_p$ are respectively the particle mass and relaxation time, and $\tilde{u}$ and $v_p$ are respectively the undisturbed fluid velocity evaluated at the particle location and the particle velocity. In two-way coupled simulations however, the undisturbed fluid velocity, $\tilde{u}$ is not directly accessible. It is common in practice to use the disturbed fluid velocity at the location of particles, $u_p$ as a replacement for $\tilde{u}$. As we shall see, this approximation can lead to significant errors in prediction of Stokes' drag when the particle size becomes on the order or larger than a fraction of the mesh size. To visually illustrate this, Figure 1 highlights the difference between $\tilde{u}$ and $u_p$.

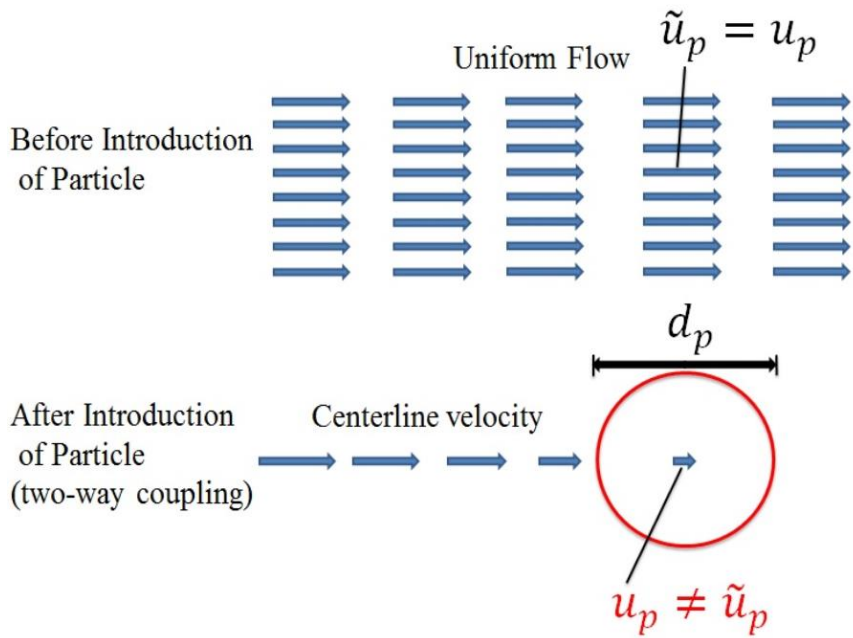

Figure 1: Illustration of how fluid velocity slows down near a particle in a two-way coupled point-particle calculation.

For small particles relative to the grid size, the difference between $u_p$ and $\tilde{u}_p$ may be small. Boivin et al. [17] suggested that the difference between $u_p$ and $\tilde{u}_p$ should be $O(d_p/dx)$, where $d_p$ is the particle diameter and $dx$ is the grid spacing. Motivated by this observation we outline the remainder of our paper.

In section two we present the point-particle method and develop error estimates for the drag force when $u_p$ is used in place of $\tilde{u}_p$. We observe a significant difference between $u_p$ and $\tilde{u}_p$ as the ratio $d_p/dx$ increases. This will have consequences not only for accurate tracking of particle trajectories but it will also lead to under prediction of the drag force and thus inaccurate prediction of the energy dissipation due to fluid-particle interactions. In section 3 the magnitude of the drag force error is estimated as a function of $d_p/dx$. To reduce contamination of statistics, in section 4 we develop a correction scheme which provides an accurate estimation of the undisturbed fluid velocity using the computed disturbed velocity near the particle. In the final section, we demonstrate the correction scheme for a particle settling in an otherwise quiescent fluid and compare that to an analytical solution. Comparison is also made with computations without the proposed correction where the disturbance fluid velocity is used in place of the undisturbed fluid velocity. The correction scheme is found to agree very well with the analytical solution while the uncorrected schemes result in very large errors. Another key conclusion from that section will be that high order interpolation schemes which are used to calculate the fluid velocity at the particle will in general perform worse than lower order schemes. This is because the higher order interpolation schemes provide a better estimate for the disturbed fluid velocity which is a worse prediction of the undisturbed fluid velocity.

Before continuing to the next section, we wish to emphasize that the purpose of this paper is not to propose a new equation of motion. For a particle placed in a turbulent flow under the most general conditions with parameters including particle size, shape, Reynolds number, volume fraction, density ratio, the equation of motion is probably more complicated than consideration of only the Stokes drag and gravity terms. The question of whether an equation of motion correctly predicts the correct physics is a validation question. We are not ready to answer that question. Rather, we wish to address the question, given an equation of motion, is the point-particle algorithm accurately computing in a situation where we know the equation of motion is valid? This is the verification question. We will show that the answer is no. The point-particle model as traditionally implemented is not verifiable for a particle moving in a low Reynolds number flow. The correction procedure therefore moves the point-particle algorithm closer to a verifiable method. The procedure we develop is a simple fix which can be incorporated with little effort into an existing point-particle code. However, we wish to emphasize that our procedure is intended to be used only in a regime where the user can argue physically that Stokes drag and perhaps gravity are the relevant terms in the particle equation of motion. In a turbulent flow, this assumption is expected to be justified for particles smaller than the Kolmogorov length scale and particle Reynolds number less than unity. The particle to fluid density ratio should be high to ensure added mass, fluid acceleration, and history terms are small in comparison to gravity and Stokes drag. In addition, the particle volume fraction should be much less than unity to ensure particle-particle interactions such as lubrication and collisions are infrequent. The particle shear Reynolds number should be much less than unity to justify neglect of the Saffman lift force [21] in favor of the Stokes drag. When the above assumptions are satisfied it is expected that the correction procedure we propose will result in more accurate calculations of the Stokes drag force which will in turn result in more accurate settling, dispersion, and fluid energy statistics.

## 2. Point-Particle Methodology

The point-particle method, strictly speaking, refers to a method of treating particles and is amenable to a host of fluid simulation methodologies including Direct Numerical Simulation (DNS), Large Eddy Simulation (LES), and Reynolds-Averaged Navier Stokes (RANS). For any hope at tackling this problem we must assume that the fluid is well-resolved in the absence of particles and that the error in the computed drag force is solely due to the improper calculation of the undisturbed fluid velocity, hence we will concern this paper only with DNS. The continuity and momentum equations used in this work are given in (1) and (2) respectively:

(1) $$\frac{\partial \rho_f}{\partial t} + \frac{\partial}{\partial x_j} \rho_f u_j = 0$$

(2) $$\frac{\partial}{\partial t}\rho_f u_i + \frac{\partial}{\partial x_j}\rho_f u_j u_i = -\frac{\partial p}{\partial x_i} + \mu \frac{\partial^2 u_i}{\partial x_j \partial x_j} + F_{g,i} - \frac{1}{V}\sum_{k}^{N_p} F_d^k\, P\{\delta(x_i - x_i^k)\}$$

Here, $\rho_f$ is the fluid density, $p$ is the pressure, $\mu$ is the dynamic viscosity, and $u_i$ is the fluid velocity. $F_{g,i} = g_i(\rho_f - \overline{\rho_f})$ is the gravitational body force per unit volume, $g_i$ is the acceleration due to gravity, $\overline{\rho_f}$ is the mean fluid density and $F_d^k = 3\pi\mu d_p(\tilde{u}_p^k - v_p^k)$ is the Stokes drag, where $\tilde{u}_p^k$ is the undisturbed fluid velocity evaluated at the location of the $k^{th}$ particle, and $v_p^k$ is the velocity of the $k^{th}$ particle. Here $P\{\delta(x_i - x_i^k)\}$ represents the numerical projection of the Dirac delta onto a computational grid. Note that (1) and (2) are written in the limit of small dispersed phase volume fraction. We solve the momentum equation (2) using a 2nd order finite difference method on a staggered mesh subject to the constraint of incompressibility. For a uniform grid of spacing $dx$, $V = dx^3$. The resulting Poisson equation for pressure is solved using a Fast Fourier transform combined with a tridiagonal matrix solver. Time stepping of fluid and particle equations is accomplished using an explicit 4th order Runge-Kutta scheme. The Lagrangian equations for the $k^{th}$ particle are given in (3) and (4):

(3) $$\frac{dx_i^k}{dt} = v_i^k$$

(4) $$m_p \frac{dv_i^k}{dt} = (m_p - m_f)g_i + F_d^k$$

Here, $x_i^k$ is the location of the $k^{th}$ particle, $v_i^k$ is the particle velocity, and $m_p$ and $m_f$ are respectively the particle and fluid mass. Dividing (4) by the particle mass yields:

$$\frac{dv_i^k}{dt} = (1 - \rho_f/\rho_p)g_i + (\tilde{u}_p^k - v_p^k)/\tau_p \tag{5}$$

The particle response time $\tau_p = (\rho_p/\rho_f)d_p^2/18\nu_f$, where $\nu_f = \mu/\rho_f$ is the kinematic viscosity.

The explicit coupling between the fluid and particle phases is now evident in equations (2) and (5). To close the point-particle algorithm we must address the calculation of $\tilde{u}_p^k$ and the projection operator $P\{\}$. As we will show in the next section, the calculation of $\tilde{u}_p^k$ generally cannot be treated separately from the specification of $P\{\}$.

In one-way coupled flows, the issue of calculating $\tilde{u}_p$ is simpler. Since fluid dynamics only influence trajectories of particles via the drag force term in (4), and particles do not feedback momentum onto the fluid, the fluid field in the neighborhood of a particle experiences no contamination owing to a single particle seeing its own disturbance flow. The only issue that arises then is that particle data may exist anywhere in space while fluid information will exist only at discrete Eulerian points. Therefore, whereas the determination of $\tilde{u}_p$ is an *estimation* problem in a two-way coupled flow, it is an *interpolation* problem in a one-way coupled flow.

Yeung and Pope [22] performed numerical simulations of turbulence and tested different interpolation schemes for tracking Lagrangian scalars. 2nd order trilinear interpolation was found to perform poorly for accurate determination of particle displacement in time while a third-order Taylor Series scheme "TS13" and 4th order cubic splines performed satisfactorily compared with exact spectral interpolation. Balachandar and Maxey [24] performed a similar study comparing trilinear, 6th order Lagrange, 2D Hermitian combined with 1D Fourier, and TS13 schemes. The partial Hermite and Lagrange schemes were found to be the most accurate while trilinear was the least accurate for determination of the Lagriangian velocity evaluated at particle locations. There does not seem to be a consensus in the community with regard to which interpolation scheme is best for tracking particles, but higher-order methods are favored. One-way coupled simulations by Calzavarini 2009 [7] and Rouson & Eaton 2001 [6] used trilinear (2nd order) interpolation; the same interpolation scheme was used by Squires & Eaton 1990, 1991 [14], [15] in two-way coupled homogeneous isotropic turbulence. 4th order Hermite interpolation was used by Elghobashi and Truesdell 1993 [26] and Ferrante and Elghobashi [3] in two-way coupled simulations and by Wang and Maxey 1993 [16] in one-way coupled simulations of particles in turbulence. Other schemes used include 4th order Lagrange interpolation by Boivin et al. [17] as well as by Sundaram and Collins 1996 and 1999 [28], [29] and 8th order Lagrange used by Ray and Collins 2011 [5].

It is important to recognize that the interpolated fluid velocity used for tracking particles in two-way coupled simulations is not the undisturbed fluid velocity found in Stokes's formula. For particles much smaller the grid spacing, the error incurred by use of the incorrect velocity may be small. Boivin et al. [17] recognized this and suggested the error would be $O(d_p/dx)$. In later sections we will verify this prediction. The ratio $d_p/dx = \Lambda$ will end up being an important physical parameter.

Considering a simulation of particle-laden homogeneous and isotropic turbulence, the main parameters are the box size which for a desired $R_\lambda$ allow the two-point correlations to die off sufficiently to not be influenced by the periodic boundary conditions. The grid spacing is chosen to resolve the Kolmogorov scale. Suppose a user decided to use a finer grid. The fluid properties would remain unchanged. For a given dispersed phase material with diameter $d_p$, $\Lambda$ increases as the grid is refined. This is equivalent to saying the source term in (2) becomes singular as the grid is refined. In the context of channel flows, a non-uniform grid is often used to resolve the inner boundary layer close to the wall. In such a scenario, whereas $\Lambda$ may be small near the channel center-plane, it may be large near the wall. Therefore, an uncorrected point-particle representation of the Stokes drag can lead to significant errors in prediction of disperse phase behavior and momentum exchanges between the two phases.

In surveying the literature, a number of physical studies used Stokes drag in two-way coupled simulations of turbulence laden by particles whose diameter was at least 30% of the unladen Kolmogorov scale [3, 19, 26, 29]. Regardless of the validity of Stokes drag for these cases, it will be important to test the accuracy in implementing Stokes drag in scenarios where the particle size becomes similar to the grid spacing, that is $\Lambda = O(1)$.

To complete the point-particle algorithm, the projection operator $P\{\}$ must be defined. Sundaram and Collins 1996 [28] derived a total energy equation for the fluid and particles under the assumption that particles are point-particles obeying Stokes drag. Because particles are not fully resolved in this context, the energy equation contains an additional dissipation term which is classified as dissipation that occurs at particle surfaces. This term is essentially a model for the contribution to true fluid dissipation that is not resolved in particle boundary layers. Sundaram and Collins showed that the numerical implementation of a point-particle algorithm would only be consistent if the weights used in the interpolation stencil to calculate the (disturbed) fluid velocity at the particle location was the same as the weights used to project the resulting drag force back onto the surrounding grid nodes. When the weights are different, the discrete energy equation evidently contains an additional error dissipation term of order one higher than the lowest order stencil used in either the interpolation or projection schemes. Therefore, in practice it is common to use the same stencil for projection and interpolation schemes. It is worth emphasizing however that the dissipation at particle surfaces is a term that only exists in a point-particle context and not in a fully resolved context. Because particle boundary layers are not resolved in the former case, this means the true fluid dissipation rate is not accurately calculated. Rather, since point-particle simulations will likely never be able to capture the true fluid dissipation rate correctly, point-particle algorithms should be tested such that the sum of the true fluid dissipation rate and the point-particle dissipation resulting from the formulation of the particle drag, along with its numerical implementation (interpolation + projection) equals the true fluid dissipation rate as calculated in a fully resolved simulation, at least in an ensemble sense. That question remains to be addressed once a properly verified and validated particle drag formulation is found.

## 3. Error Estimation

In this section, we develop estimates for the error in calculation of Stokes drag when the disturbed fluid velocity is used in place of the undisturbed fluid velocity. The model problem we consider will test the accuracy of the point particle model in a regime where the Stokes formula is known to be accurate, namely, uniform flow passed a stationary sphere at low Reynolds number. This is the scenario depicted in Figure 1. Prior to introduction of the particle, the fluid velocity will be uniform everywhere, and the undisturbed fluid velocity at the location of the particle will be equal to the inflow velocity. When the particle is introduced, the particle slows down the fluid velocity locally owing to two-way coupling. The computational implementation is depicted in Figure 2. A point-particle is initialized at a fluid cell-face and held fixed. This formulation ensures there will be no interpolation error. The fluid velocity at the location of the particle is precisely the fluid velocity on the cell face upon which the particle rests. The projection of the drag force onto the fluid will reduce the fluid velocity near the particle. The particle will then sample a lower velocity than the undisturbed value and will project a new force onto the fluid grid. Viscous diffusion will tend to regularize the influence of the projected force.

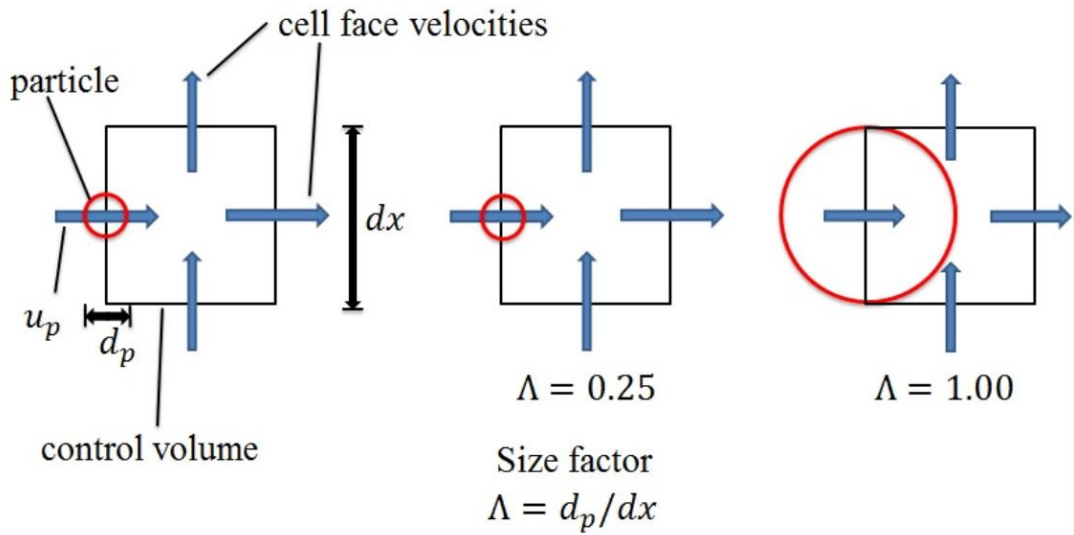

Figure 2: Model Problem for error estimation.

We therefore should expect an equilibrium to result with a region of slowed fluid near the particle and a drag force calculated as $F_d = 3\pi\mu d_p(u_p - v_p) < 3\pi\mu d_p(\tilde{u}_p - v_p)$. Note here that $v_p = 0$ since the particle is stationary. Three-dimensional simulations are performed with periodic boundary conditions in lateral and transverse directions and inflow-outflow boundary conditions in the streamwise direction. Because the Stokes disturbance field decays very slow away from the particle $\sim 1/r$, a large box was necessary to ensure the periodic boundary conditions did not influence the calculation of the drag force. In these test cases and for the remainder of the paper, the number of grid points $N^3 = 128^3$ was found to yield accurate and converged results. The grid spacing is equal in all directions, $dx = dy = dz$. The particle Reynolds number is $Re_p = \tilde{u}_p d_p / \nu_f = 0.1$.

We also examine the effect of the projection scheme on the error. In the symmetric formulation which uses the same weights for interpolation and projection, all of the computed drag force is projected onto one grid point. This means the disturbance created by the point source is larger than if it was spread over many grid points. Capecelatro and Desjardins [10] noted that the source term is singular under grid refinement and proposed a two-step procedure. The first step is to mollify or project the calculated drag force onto neighboring grid points using a Gaussian stencil of compact support. For particles of size comparable to the grid spacing, the mollification still results in a singular source term as the grid is refined. Capecelatro and Desjardins proposed a second step whereby the mollified drag force is diffused to surrounding grid points. The second step is performed for particle sizes $\Lambda > 1/3$ and was demonstrated to result in a converged projection filter that did not depend on the grid spacing. To understand the role of using a Gaussian projection scheme, we have implemented the mollification procedure used in that work for a filter width comparable to the mesh spacing (the cutoff of the filter is $r_{cutoff} > 1.5 dx$). The Gaussian filter is normalized such that the sum of weights assigned to all grid points is unity.

Error in steady state drag for the one point and Gaussian projection schemes for different particle sizes are summarized in Figure 3. The error is seen to increase with the particle to mesh size and is of the $O(\Lambda)$, consistent with Boivin et al.'s [17] prediction. The error was found to be relatively insensitive to Reynolds number in the low Reynolds number limit. Note that the error is non-negligible even for relatively small $\Lambda = 0.1$. In all cases the Gaussian stencil reduces the computed drag force error, but not significantly. Incorporation of the diffusion procedure may reduce error incurred for the $\Lambda = 0.5$ and $\Lambda = 1.00$ cases, however it is still clear from these results that convergence of the Gaussian filter width does not imply convergence of $F_d$ to the Stokes drag solution. In principle, a projection of infinite support, or in practice, using a Gaussian mollification that incorporates a large number of grid points would reduce the error in calculated drag force. This is because the influence of the calculated force distributed among many grid points would result in a smaller disturbance at the location of the particle. However, communication with a large number of grid points would become impractical for simulations tracking millions of particles.

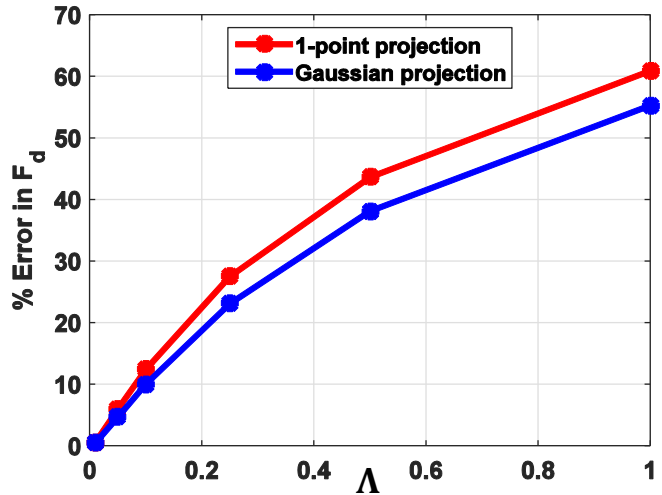

Figure 3: Percent error in calculation of $F_d$ for flow passed a stationary point particle.

It is important to note that these are simply error estimates. As a particle moves through the grid, it will sample velocities at different distances from its center via the chosen interpolation scheme. The projection scheme will distribute the calculated force over several grid points so the error in drag force as computed when a particle moves in an arbitrary direction through the grid will be different from what we computed for the special case that a particle is collocated with a flow velocity mesh point. It is also important to note that these are steady errors obtained by integrating the fluid equations for approximately one hundred viscous relaxation times. (This was necessary to ensure a steady state solution since the velocity disturbance solution to the unsteady Stokes equation decays slowly as $\sim \exp(-r^2/\nu_f t)$ , Basset p.288 [1].) Therefore, we may expect that the error in drag force for a moving particle will depend on the particle Stokes number, $St = \tau_p/\tau_{visc}$. This question will be explored in section 5. The results in this section however demonstrate the order of the error incurred by use of the disturbed fluid velocity in the Stokes drag formula. While different projection schemes may reduce this error, they do not explicitly address the problem of incorrectly using $u_p$ in the drag expression compared with $\tilde{u}_p$.

## 4. Proposed Correction Method

In the previous section, we observed that simply spreading the computed drag force over more fluid grid points will not significantly reduce the error in the calculated force. At initialization, a particle has not created its own disturbance field, so the computed drag force will be correct. However, as time progresses, the region of slowed flow grows around the particle. The centerline velocity for low Reynolds number flow passed a stationary point-particle is depicted in Figure 4. Near the particle (and within a few grid points of its center), the fluid velocity is considerably slower than the undisturbed fluid velocity. In this scenario, the fluid velocity far from the particle is equivalent to the undisturbed velocity at the particle location. In addition, the streamwise fluid velocity is nearly symmetric about the center of the particle. Motivated by these observations, we may propose a correction scheme of the form:

$$\tilde{u}_p \approx u_p + C_1 dx \frac{\partial u_p}{\partial x} + C_2 dx^2 \nabla^2 u_p + \cdots \tag{6}$$

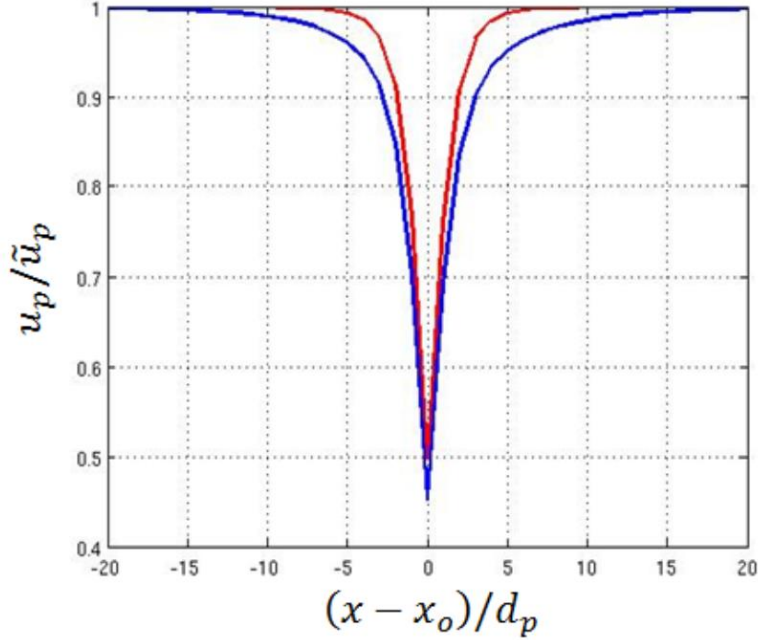

Figure 4: centerline velocity for flow passed a stationary point particle at different times, $t^* = \frac{t v_f}{d_p^2} \approx 1\ (red),\ 24\ (blue)$, $N^3 = 128^3$, $\Lambda = 1$, Gaussian projection.

Since the disturbance flow is nearly symmetric, the first derivative is very small at the particle location. This allows us to simplify (6) to:

$$\tilde{u}_p \approx u_p + C(\Lambda, x_i) dx^2 \nabla^2 u_p \tag{7}$$

Equation (7) is the central statement in our proposed scheme. The proposition is that a good approximation for the undisturbed fluid velocity is a truncated power series expansion which incorporates the disturbance fluid velocity and adds a correction owing to the disturbance field each particle has generated for itself. Note, this is not the Faxen [33] contribution which is a correction to the drag formula which accounts for the fact that in a multiple particle system, particles see a disturbance field created by other particles. In this context, we are seeking only to accurately calculate the Stokes drag. In the Stokes formula, a particle does not see its own disturbance field so it is appropriate to remove that disturbance which a point-particle creates for itself. Such an estimate is expected to be good when the particle volume fraction is small such that particle-particle interactions are not a leading order effect in momentum or energy coupling.

In equation (7), $C(\Lambda, x_i)$ is still unknown. As we will show, this coefficient will depend on both the particle size relative to the grid spacing $\Lambda$, and in general will depend on the location of a particle within a grid cell. Assuming this $C$-field can be found as a function of $\Lambda$, then the problem of *estimating* $\tilde{u}_p$ has been transformed to a problem of *interpolating* $C$. The latter is a simpler problem to solve. As we will see, the reason interpolation of $C$ will be successful whereas interpolation of the fluid velocity field is unsuccessful is that the $C$-field will involve a set of coefficients that are independent of time, i.e. there is no feedback on the $C$-field. Interpolation of $C$ will be successful because it represents uncontaminated data whereas the fluid velocity field is contaminated by the particle disturbance field and hence is not directly suitable for interpolation.

$C$ will be constructed by reverse engineering. We will hold a particle fixed in an otherwise uniform flow and ask, what value of $C$ for a given $\Lambda$ makes (7) an equality? The drag force appearing in the momentum equation (2) will be computed exactly as $F_d = 3\pi\mu_f d_p \tilde{u}_p$. Since $\tilde{u}_p$ is identical to the prescribed inflow velocity, it easy to calculate in steady-state the value of $C$ such that

(8) $$C = (\tilde{u}_p - u_p)/dx^2 \nabla^2 u_p = (u_{BC} - u_p)/dx^2 \nabla^2 u_p$$

This procedure is done for each $\Lambda \in [0.01, 0.05, 0.1, 0.25, 0.5, 1.0]$. We choose $Re_p = 0.1$ for each of these cases, with results having little dependence on $Re_p$ in the low $Re_p$ limit. Since at low Reynolds number, the momentum transport equation is almost linear and quasi-steady (inertial effects are negligible near the particle), it is appropriate to integrate the fluid equations to steady state and take the resulting value of $C$. We will show in section 5 that treating $C$ as time invariant works well even for a moving particle. In Table 1, we present a grid study of the computed value of $C$ for $\Lambda = 1$ for a particle collocated on the velocity point, and present convergence to about the third digit for the $128^3$ grid. Two viscous CFL numbers, both well below the Runge-Kutta stability boundary were explored and found to have little effect on the computed value of $C$. So far, we have demonstrated how to calculate exactly the drag force the fluid experiences owing to a particle placed and held fixed on a fluid cell-face.

| Grid | $CFL_{visc} = 0.36$ | $CFL_{visc} = 0.1$ |
|---|---|---|
| $8^3$ | 0.227316 | 0.227311 |
| $16^3$ | 0.238328 | 0.238325 |
| $32^3$ | 0.244916 | 0.244956 |
| $64^3$ | 0.247345 | 0.247346 |
| $128^3$ | 0.247746 | 0.247747 |
| Table 1: Grid study for $C$, $\Lambda = 1$. | | |

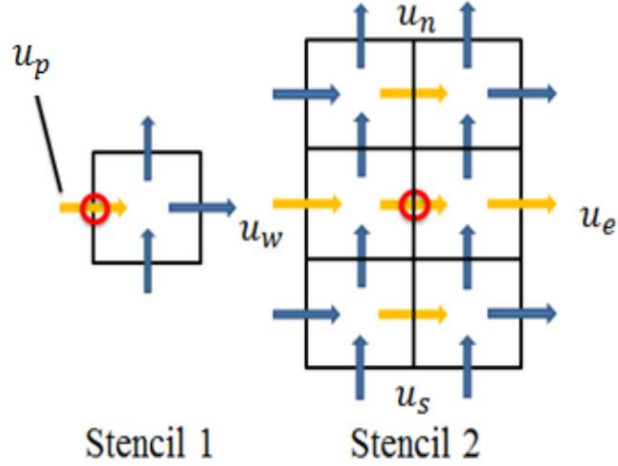

Figure 5: Two-dimensional stencil for flow passed stationary point-particle.

This scenario in two and three dimensions are shown respectively in Figures 5 and 6. The difference between a traditional interpolation scheme and the proposed correction is summarized in equations (9), (10a), and (10b):

(9) $Stencil\ 1\ (uncorrected)$: $\qquad \tilde{u}_p = u_p$

(10a) $Stencil\ 2\ (2D)$: $\tilde{u}_p = u_p + C dx^2 \nabla^2 u_p$
$\qquad\qquad\qquad\qquad = (1 - 4C) \times u_p + C(u_n + u_s + u_e + u_w)$

(10b) $Stencil\ 2\ (3D)$: $\tilde{u}_p = u_p + C dx^2 \nabla^2 u_p$
$\qquad\qquad\qquad\qquad = (1 - 6C) \times u_p + C(u_n + u_s + u_e + u_w + u_t + u_b)$

Note all results are implemented with the 3D stencil; the 2D stencil is shown only for illustration. It is apparent that stencil 2 can be regarded as a non-convex interpolation scheme with a mixture of positive and negative interpolation weights. Since the fluid velocity away from the particle is larger than the velocity near the particle, a non-convex scheme can estimate a velocity $\tilde{u}_p$ which is greater than all velocity values used $u_p, u_n, u_s, u_e, u_w, u_t, u_b$. Note this scheme is conservative since in the absence of a disturbance flow, the scheme would correctly predict $\tilde{u}_p = u_p$.

In general, a particle will not be located at a grid node (fluid cell-face in a staggered mesh configuration). Therefore the proposed scheme must be extended to the arbitrary position of a particle in a grid cell. Unfortunately, the $C$ value we have obtained for a particle located at a cell face which we will call $C_1$, will not work for an arbitrary particle location.

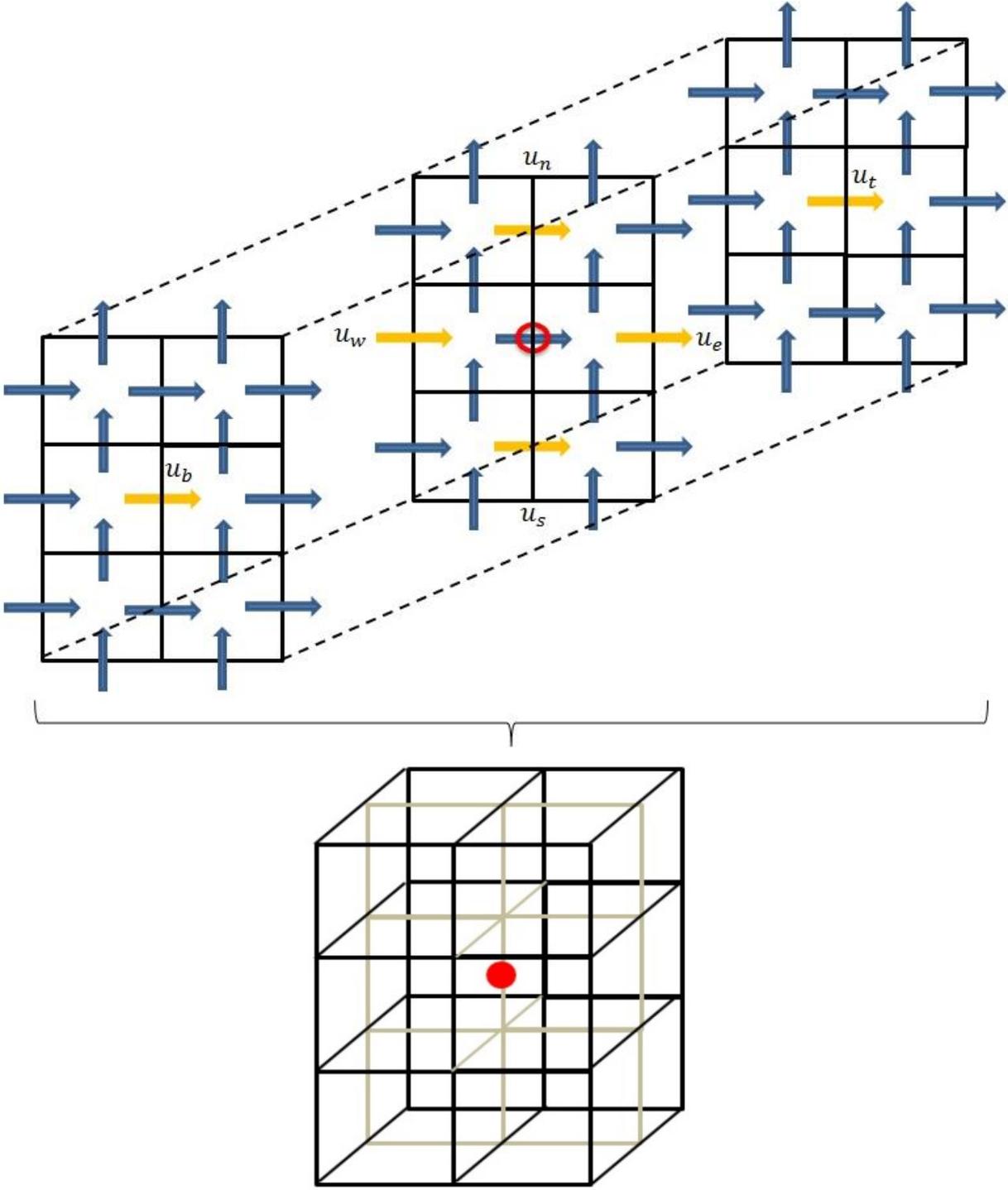

Figure 6: Three-dimensional stencil for flow passed stationary point-particle, (orthogonal velocity components not shown for clarity).

Based on the previous observations we propose a solution to the arbitrary particle location by solving for $C$ at a few points on, and within a grid cell. This scenario is shown in Figure 7. We consider a decomposition of a cubical grid cell into eight cubes of side length one half ($dx/2$) the physical grid spacing, $dx$. Nodes of the original cube (shown in blue) correspond to fluid velocity cell-faces as depicted in Figure 8. Hence an appropriate value of $C$ for that location is $C_1$. $C$ values can then be obtained at the vertices of a single smaller cube by placing a particle at each of those locations and solving for the resulting C values via (8).

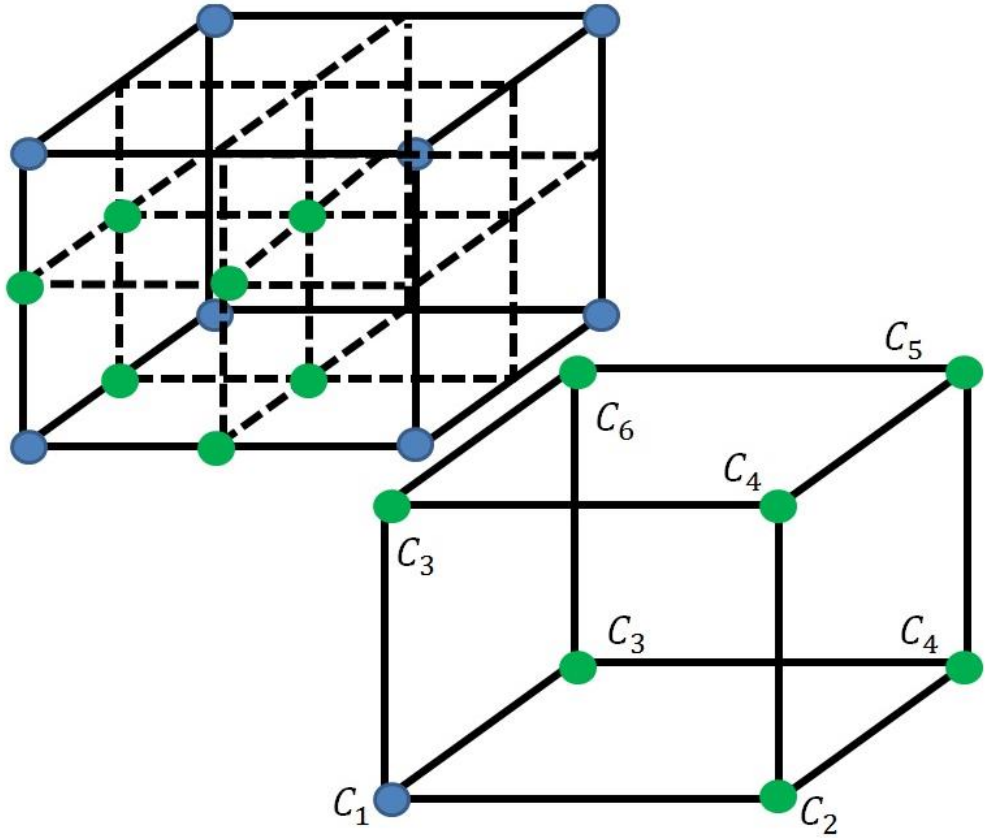

Figure 7: Formulation of C field in uniform grid.

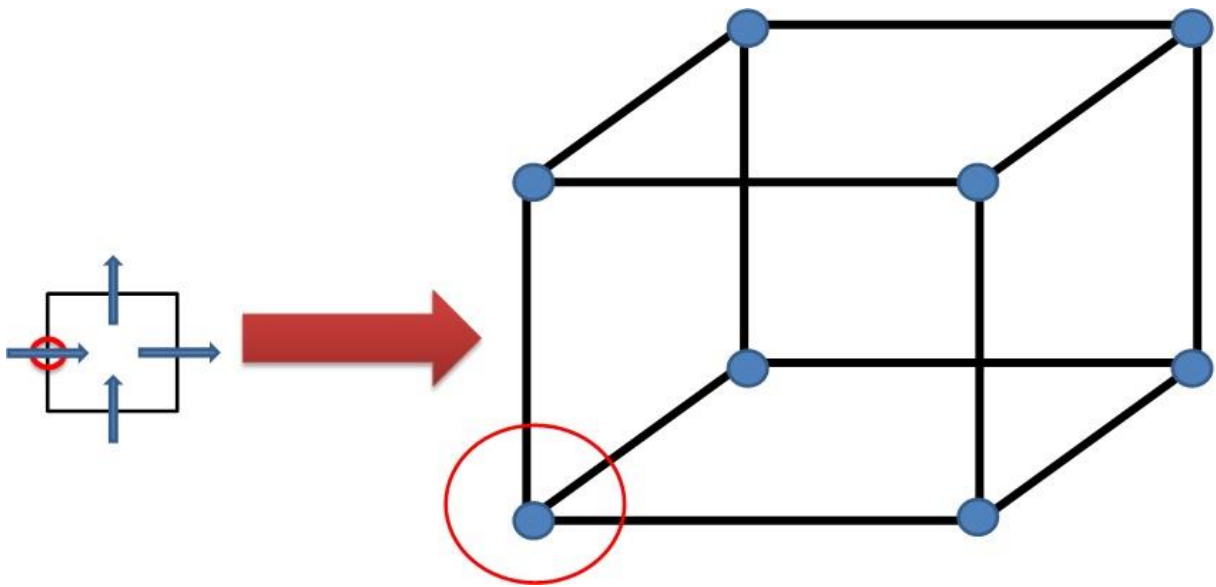

Figure 8: cell-faces correspond to cube vertices.

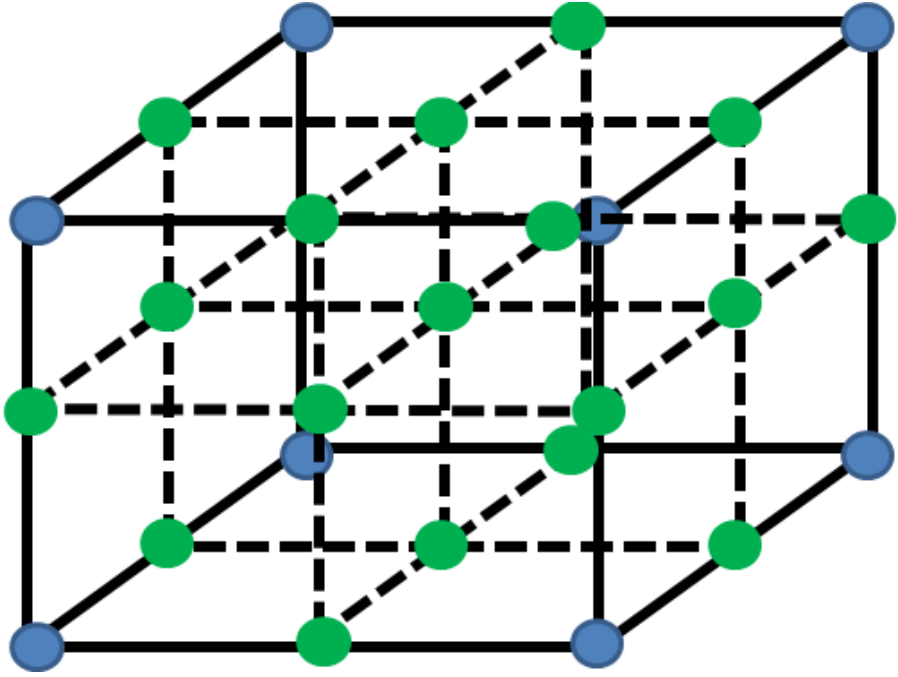

Figure 9: Scalar field $C$ embedded in a grid cell; blue vertices indicate cell faces.

Owing to symmetry of the Stokes solution, six values of $C$ uniquely define the eight vertices. Further, within a grid cell, there is no distinction among a particle being in one of the smaller cubes over another. Therefore, determination of eight (six unique) $C$ coefficients in one small cube determines the $C$ coefficients for the nodes of all eight cubes. As depicted in Figures 9 and 11, this procedure will result in a periodic scalar field $C$ embedded in each grid cell. In Table 2, we tabulate the six unique coefficients for different particle to grid sizes. For a chosen $\Lambda$, a set of six numbers can be used in conjunction with (7) to obtain a good estimate for the undisturbed fluid velocity at the particle location. Since the $C$-field has no feedback on it, the tabulated coefficients are not contaminated by a disturbance and hence independent of time. As we will show in the next section for a particle moving through the grid in an arbitrary direction, a low memory access trilinear interpolation method for $C$ will produce accurate results for all of the particle sizes we consider. For particle sizes that are different than those presented here, it is straightforward to estimate the $C$ coefficients by interpolating the $C$ data using smooth cubic splines. A plot of the $C$ data and the splines is shown in Figure 10. The spline equations used for this interpolation can be found in the Appendix.

| $\Lambda$ | $C_1$ | $C_2$ | $C_3$ | $C_4$ | $C_5$ | $C_6$ |
|---|---|---|---|---|---|---|
| 1.0 | 0.248 | 0.312 | 0.340 | 0.440 | 0.681 | 0.518 |
| 0.5 | 0.246 | 0.309 | 0.337 | 0.435 | 0.671 | 0.512 |
| 0.25 | 0.242 | 0.303 | 0.329 | 0.422 | 0.644 | 0.493 |
| 0.1 | 0.231 | 0.285 | 0.306 | 0.389 | 0.579 | 0.447 |
| 0.05 | 0.218 | 0.267 | 0.282 | 0.356 | 0.522 | 0.403 |
| 0.01 | 0.190 | 0.236 | 0.239 | 0.311 | 0.460 | 0.335 |
| Table 2: $C$-field coefficients for different values of $\Lambda$. | | | | | | |

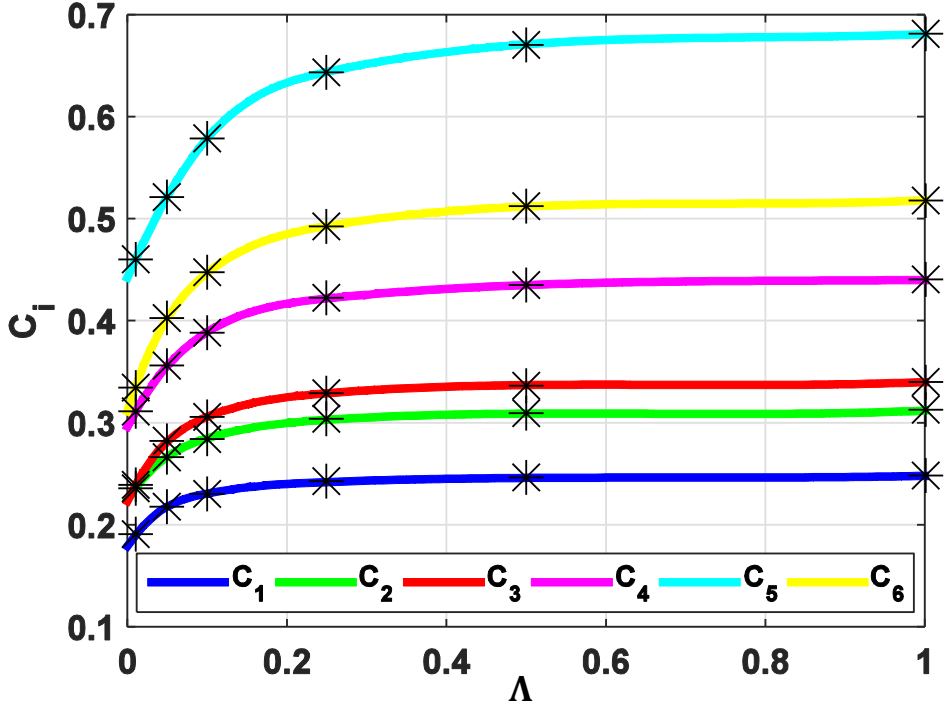

Figure 10: $C_{1-6}$ vs. $\Lambda$ and interpolating cubic splines.

The $C$ data is smooth over the range of $\Lambda$ we considered. As shown in Figure 10, $C_i$ is more sensitive to $\Lambda$ for smaller $\Lambda$. However errors in estimating $C_i$ for small $\Lambda$ would not be significant since the disturbance velocities generated by point particles at small $\Lambda$ are relatively small. For particle sizes $\Lambda \geq 0.25$, the $C_i$ field shows little variation. Therefore, an accurate approximation for these $C$ coefficients can be taken by evaluating the spline interpolants for a given $\Lambda$ chosen for a particular study.

While the $C$ coefficients were obtained for a particle held fixed in a uniform flow parallel to the x-direction, the same $C$ coefficients will apply to the orthogonal directions. Owing to the orthogonality of the three velocity components (for a Cartesian mesh) in a staggered formulation, the $C$ field ends up being transposed for the v and $w$ components, compared with the $u$ component. This is depicted in Figure 11. Notice that the $C$ field is periodic so that it can be completely characterized by two layers. The blue nodes represent cell-faces in the respective $u$, v, and $w$ momentum equations which are consistent with the $C_1$ value while green nodes represent particle locations within a grid cell. Since the $C$ field is periodic, only 6 constants are needed which are the same for all grid cells for all momentum directions (up to a rotation of the coordinate frame). Therefore this method requires very little storage and can be easily integrated into an existing finite difference code.

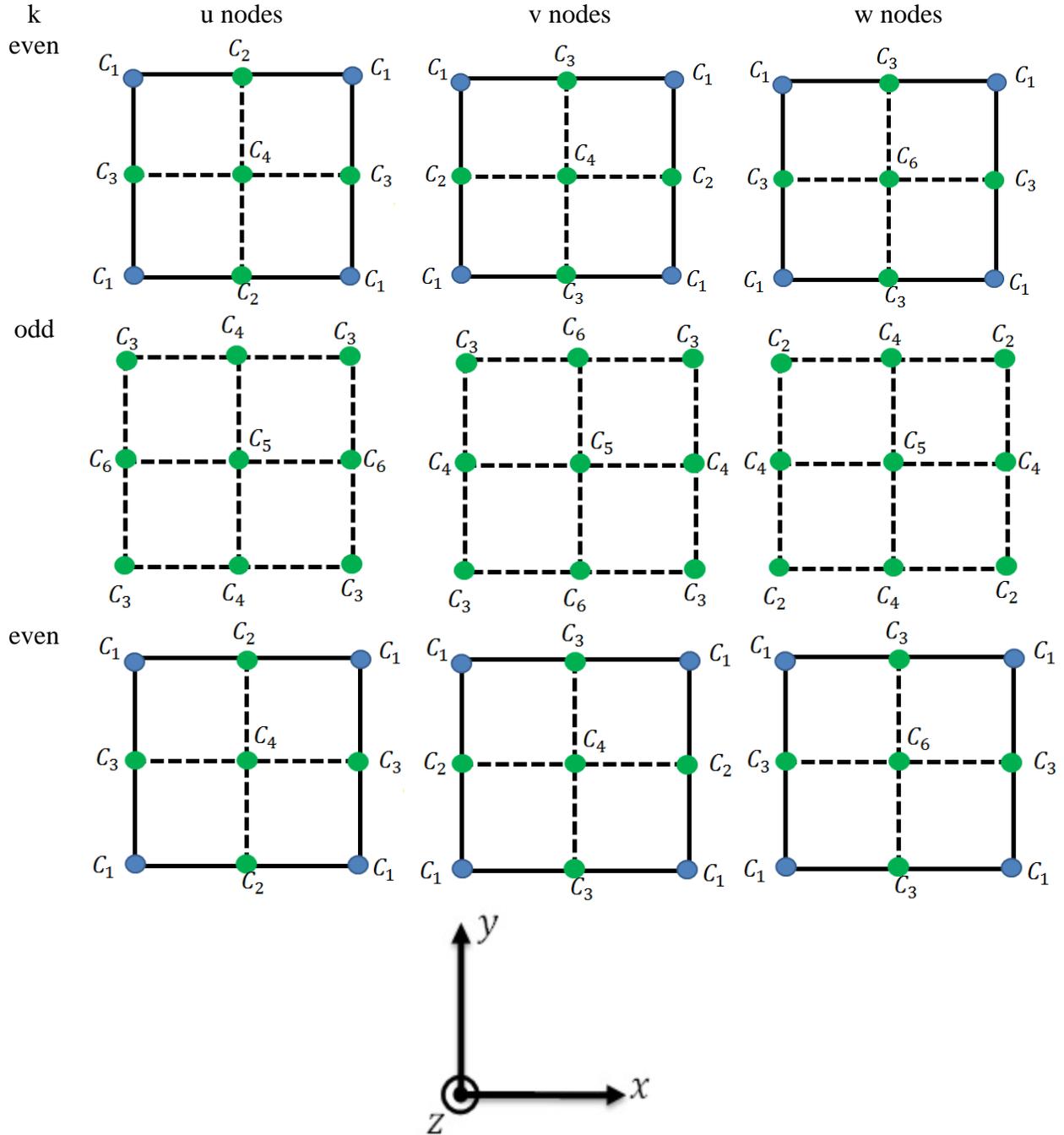

Figure 11: Distribution of $C$ field within a cell for $u, v, w$-velocity components. The blue node for each case represents the location where the corresponding velocity component is stored.

### 4.1 Algorithm Summary: to calculate undisturbed velocity for arbitrary particle location

We now summarize the proposed algorithm. For an explicit scheme, we will have fluid $u_i^n$ and particle $v_i^n$ velocity data at the $n^{th}$ time step and we wish to advance to the next timestep, $(n + 1)$. Here we will only discuss the force calculation as this is the only calculation that differs from a classical Euler-Lagrange method. To calculate $F_{d,i}$, which will be needed to update both particle and fluid velocities, we recognize the $i^{th}$ component of the drag force $F_{d,i} = 3\pi\mu d_p(\tilde{u}^n{}_{p,i} - v_i^n)$, where $\tilde{u}^n{}_{p,i}$ is the undisturbed fluid velocity evaluated at the particle location. Note that this undisturbed velocity need not be constant in time if for example a particle is in a time evolving flow (turbulence e.g.). Therefore, we wish to calculate the $i^{th}$ component of the undisturbed fluid velocity at time step $n$ via $\tilde{u}^n{}_{p,i} \approx u^n_{p,i} + C(\Lambda, x_i)dx^2\nabla^2 u^n_{p,i}$, where the position $p$ within a cell may be arbitrary.

The algorithm steps are:

- 1. $C_{1-6}$ are prescribed at the beginning of a simulation for the chosen particle size $\Lambda$ (multiple $C$ arrays could be used if the particle field is polydisperse).
- 2. At time step $n$, for a given particle location in the Eulerian grid, determine its position with respect to one of the eight small cubes within the physical grid cell. Note that this location may be different for the $u$, v, and $w$ components owing to the staggered grid and transpose of the C field. However this step can be accomplished very easily with a few if statements.
- 3. Using trilinear interpolation, estimate the value of $C$ at the particle location using the vertices of the small cube. Again here, the interpolated value of $C$ will be different for the $u$, v, and $w$ directions.
- 4. Using trilinear interpolation, calculate $\tilde{u}^n_{p,i} \approx u^n_{p,i} + C_{int} dx^2 \nabla^2 u^n_{p,i}$ where $u^n_{p,i}$ and $\nabla^2 u^n_{p,i}$ are interpolated from the cell-faces of the grid. For a 2$^{nd}$ order scheme, a consistent discretization of $\nabla^2 u$ would use a 2$^{nd}$ order central scheme. Note $\nabla^2 u$ is a quantity that will already exist at fluid cell-faces since it is needed to update the momentum equations even in the absence of this correction.
- 5. With the estimate of $\tilde{u}^n_{p,i}$, the particle forcing term $3\pi\mu_f d_p(\tilde{u}_p - v_p)$ may be formed. The particle equations are ready to be updated using the appropriate time advancement scheme.
- 6. To update the fluid equations, it is necessary to project the computed force back to the fluid grid. The coefficients $C_i$ we have reported are consistent with using trilinear weights (based on the velocity cell face distances) to project the computed drag force back to the fluid cell faces.

It is important to re-emphasize here that the reported $C$ coefficients are consistent with using trilinear interpolation in steps 4 and 6. In step 3, any interpolation scheme is appropriate. This is not a drawback of the method. The ability to correct for the disturbance flow using low memory access methods (trilinear) compared with higher order interpolation schemes makes this method advantageous from a runtime prospective. From an implementation prospective, the only modifications to an existing Euler-Lagrange code would be: (1) addition of a trilinear interpolation routine (if one does not already exist in the code), (2) a simple set of conditionals to classify a particle as being within one of eight 8 smaller cubes within a fluid grid cell, (3) either loading Table 2 directly into the code or having a preprocessor that evaluates the splines for a given Λ and loads those 6 coefficients into the code, and (4) finally a summing step that forms the suitable linear combination of interpolated fluid velocity, $C_{int}$, and Laplacian components. All of these pieces can be incorporated into an existing code with relative ease.

## 5. Method Verification

We have developed a correction procedure to calculate the undisturbed fluid velocity at the location of a particle placed and held fixed in an arbitrary location in the fluid grid. However, we must ensure the proposed procedure is capable of accurately capturing the dynamics of a moving particle, since in most applications a particle is free to move owing to the stresses the fluid exerts on it. An appropriate verification problem is that of particle settling in an unbounded and otherwise quiescent fluid subject to gravity. The problem setup is shown in Figure 12.

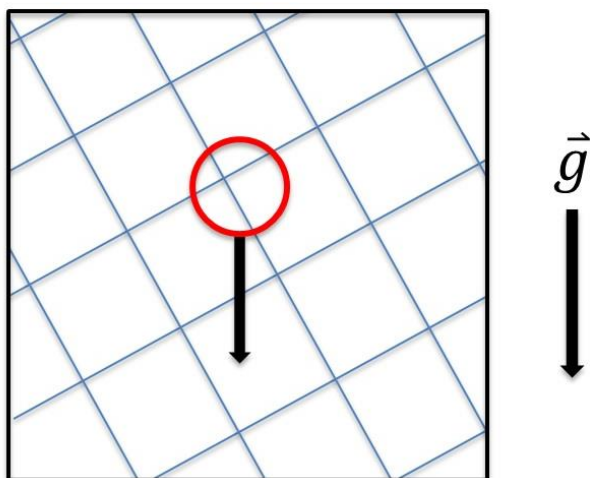

Figure 12: Particle settling in an unbounded fluid where the gravity vector is chosen at an arbitrary tilt with respect to the grid.

We consider a three-dimensional box with $128^3$ grid points and periodic boundary conditions on all sides. To test the robustness of the scheme, we do not align the gravity vector with the grid. We have chosen the gravity unit vector to be $\hat{g} = (1, (1+\sqrt{5})/2, e)/|g|$. This ensures the particle achieves good sampling of the fluid field as it passes within and from grid cell to grid cell. The particular choice of the gravity vector is arbitrary, but specifying $\hat{g}$ such that the ratio of any two components of $\hat{g}$, that is, $\hat{g}_x/\hat{g}_y$, $\hat{g}_x/\hat{g}_z$, or $\hat{g}_z/\hat{g}_y$, is an irrational number, is not arbitrary. Such combinations would ensure that the collection of points along the particle trajectory would form a dense set [4, 27] within the domain of each mesh element and thus all possible interpolation scenarios would be tested in our verification study.

With this choice of setting, while the analytical solution to the particle equation will yield a particle velocity parallel to the gravity vector, the simulated results will not, and rather have some component of velocity perpendicular to the gravity direction. This drift velocity is unphysical (for an isolated particle) and will be owing to bias in interpolation of the fluid velocity at the particle location. Finally, because this verification problem admits an analytical solution in the low Reynolds number limit, it is a robust problem to assess the accuracy of the proposed scheme in comparison with other standard techniques.

Table 3 presents the simulation runs for a particle settling under gravity for various flow conditions. Note that the proposed scheme was developed for flow passed a stationary particle ($St \rightarrow \infty$), at $Re_p = 0.1$, so it will be important to consider the effectiveness of the proposed scheme over a range of these parameters. Here, the Stokes number $St = \tau_p/\tau_{visc}$, where $\tau_p$ is the particle response time and $\tau_{visc} = \nu_f/dx^2$, where $\tau_{visc}$ is the relaxation time of the grid. Since the particles are not fully resolved, it is not appropriate to use $d_p$ as the length scale for viscous relaxation in the fluid. In a turbulent flow, the definition of $\tau_{visc} \approx \tau_\eta$ and is consistent with our choice since, $\eta = O(dx)$ in DNS [25]. We test particle sizes in the range $\Lambda \in [0.01, 1.00]$, Stokes numbers $St \in [0.25, 10]$, and particle Reynolds numbers $Re_p \in [0.05, 0.50]$. The proposed scheme is compared with two methods of interpolation, trilinear, and 4$^{th}$ order Lagrange. While there are other interpolation schemes that could be tested, some general conclusions will arise. Finally, while the Stokes drag is formally derived in the limit of $Re_p \rightarrow 0$, the deviation of the Stokes drag from experimental observations is within a few percent for particle Reynolds numbers up to unity [23]. Hence, examination of particle Reynolds numbers up to 0.5 in this study will be within the regime of validity of the Stokes drag.

| Run | $Re_p$ | $St$ | $\Lambda$ | $CFL_v$ | $CFL_p$ |
|---|---|---|---|---|---|
| | | Size Study | | | |
| A | 0.1 | 10.0 | 1.0 | 0.36 | 0.006 |
| B | 0.1 | 10.0 | 0.5 | 0.36 | 0.006 |
| C | 0.1 | 10.0 | 0.25 | 0.36 | 0.006 |
| D | 0.1 | 10.0 | 0.1 | 0.36 | 0.006 |
| E | 0.1 | 10.0 | 0.05 | 0.36 | 0.006 |
| F | 0.1 | 10.0 | 0.01 | 0.36 | 0.006 |
| | | Reynolds number Study | | | |
| G | 0.05 | 10.0 | 1.0 | 0.36 | 0.006 |
| H,A | 0.1 | 10.0 | 1.0 | 0.36 | 0.006 |
| I | 0.25 | 10.0 | 1.0 | 0.36 | 0.006 |
| J | 0.50 | 10.0 | 1.0 | 0.36 | 0.006 |
| | | Stokes number Study | | | |
| K,A | 0.1 | 10.0 | 1.0 | 0.36 | 0.006 |
| L | 0.1 | 5.0 | 1.0 | 0.36 | 0.012 |
| M | 0.1 | 1.0 | 1.0 | 0.36 | 0.06 |
| N | 0.1 | 0.5 | 1.0 | 0.36 | 0.12 |
| O | 0.1 | 0.25 | 1.0 | 0.36 | 0.24 |

Table 3: List of runs, Courant numbers: Viscous: $CFL_v = 6\nu_f \Delta t/dx^2$, Particle: $CFL_p = \Delta t/t_p$.

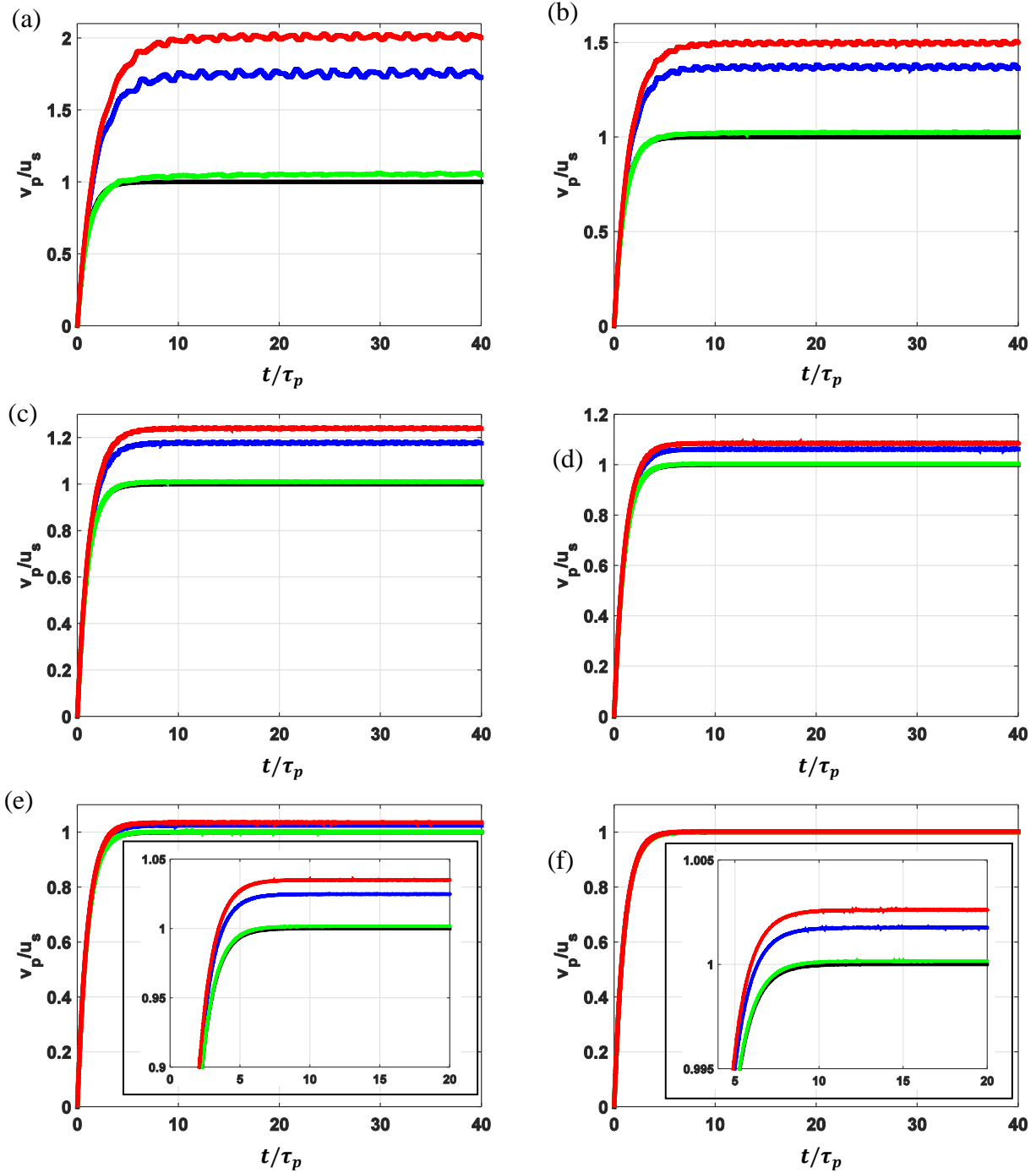

Figure 13: Particle transient settling velocity for different particle sizes, $(a)\Lambda = 1$, $(b)\Lambda = 0.5$, $(c)\Lambda = 0.25$, $(d)\Lambda = 0.1$, $(e)\Lambda = 0.05$, $(f)\Lambda = 0.01$, red-4$^{th}$ order Lagrange, blue-2$^{nd}$ order trilinear interpolation, green-correction scheme, black-analytical solution, inset shows the ordering of respective schemes' steady state behavior is preserved for all $\Lambda$.

Settling velocity histories for different sized particles are shown in Figure 13. The four curves shown compare the proposed correction scheme to particle tracking using trilinear and fourth order Lagrange. The analytical solution $u_{ss} = g\tau_p(1 - \rho_f/\rho_p)(1 - \exp(-t/\tau_p))$ is shown for comparison. Here, we are using the same weights for the uncorrected interpolation and projection schemes (trilinear, 4th order Lagrange) based on the recommendation of Sundaram and Collins 1996 [28]. The correction scheme follows the analytical solution for all particle sizes with a maximum steady state error of approximately 5% for the largest particle size, $\Lambda = 1$. Both the trilinear and Lagrange schemes significantly overshoot the steady state analytical settling velocity. For $\Lambda = 1$, the trilinear scheme overshoots the steady value by approximately 75%, while the Lagrange-4 scheme overshoots by approximately 100%. Because the interpolation schemes are measuring a fluid velocity that contains the particle disturbance flow, the measured slip velocity $u_s^{measured} = u_p - v_p$ is less than the slip velocity upon which the Stokes drag analytical depends, i.e. $u_s^{actual} = \tilde{u}_p - v_p$. Since the slip velocity is underpredicted, as a consquence, so is the drag force. As a result, lower drag means a higher settling velocity.

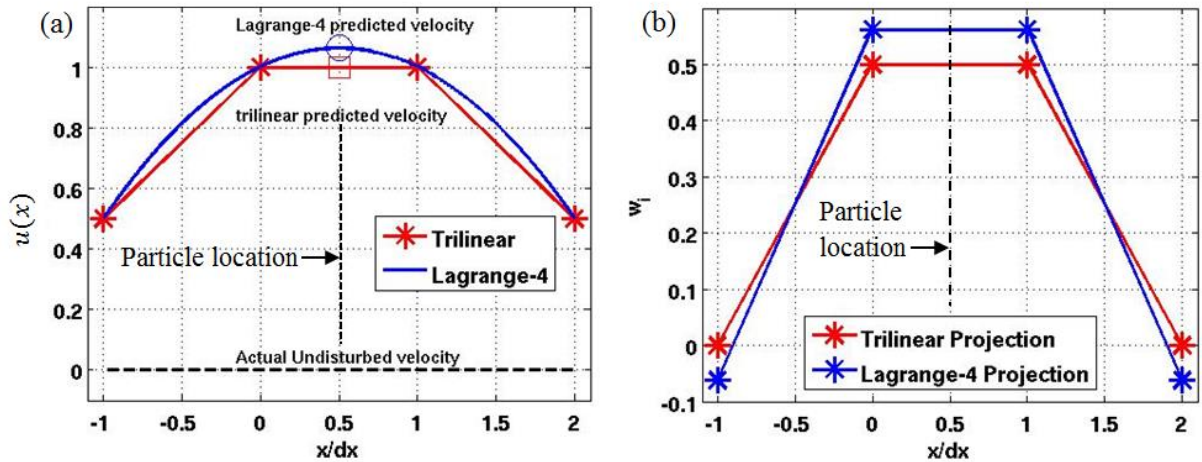

Figure 14: (a) velocity predicted by interpolation schemes compared with actual undisturbed fluid velocity, (b) spatial distribution of projection weights

At first glance, it is not intuitive why 4th order Lagrange behaves more poorly than trilinear interpolation. A simple explanation is depicted in Figure 14. Consider a particle settling under gravity. For simplicity, we consider a one-dimensional problem and a particle whose instantaneous position is halfway between two fluid cell-faces. The undisturbed fluid in this scenario is stationary. Due to gravity, the particle moves with a velocity greater than the surrounding fluid's velocity. But, owing to the momentum coupling between the particle and the fluid, the particle will tend to drag fluid with it, so that the fluid velocity near the particle is higher[2] than the undisturbed value. Since the disturbance flow is symmetric about the particle center in a Stokes flow, the fluid velocity measured at the grid points surrounding the particle will take a concave shape (the precise magnitude is not important here, only that the curvature is in the direction depicted in Figure 14. Trilinear (linear in 1D) interpolation would sample the fluid velocity from the nearest neighbor grid points and predict some fluid velocity at the particle location. The 4$^{th}$ order Lagrange scheme however is able to account for the curvature in the fluid velocity field and to recognize the fluid (disturbed) velocity at the particle location is greater than than the trilinear prediction. If the fluid velocity data were not contaminated by a disturbance flow, the 4$^{th}$ order Lagrange estimate would be better than the trilinear prediction. However, the goal is to compute $\tilde{u}$, which in the present problem is identically zero. So in fact, while the 4$^{th}$ order Lagrange scheme provides a better estimate of the disturbed velocity at the particle location, this corresponds to a worse estimate of the undisturbed velocity. Therefore we may expect in general that higher order schemes will tend to provide better estimates of the disturbed flow which is counter to the problem of estimating the undisturbed fluid velocity. We have also tested cubic splines, Lagrange 6$^{th}$, Lagrange 8$^{th}$ order schemes (not shown for clarity) which cooroborates this claim.

---

[2] Note that the notion of higher or lower depends on the reference frame of the observer.

Taking the 4$^{th}$ order Lagrange scheme as an example, the problem is exacerbated when considering what happens in the projection step. After estimating the drag force based on an interpolated fluid velocity, the drag force must be projected back to the Eulerian grid. If the same weights that were used for interpolation are also used for projection, we can see that the 4$^{th}$ order Lagrange scheme would project a larger percentage of the force on grid points close to the particle. As depicted in Figure 14 (b), the weights associated with the Lagrange-4 scheme are 0.5625 for both grid points immediately surrounding the particle and -0.0625 for the next nearest grid points. So the 4$^{th}$ order Lagrange scheme will tend to create a larger disturbance flow in the neighborhood of the particle than does the trilinear scheme. Any interpolation scheme which has larger weights near the particle and smaller away from the particle will yield similar behavior. Examples of such schemes include higher order Lagrange, Cubic Spline, and Gaussian interpolation/projection schemes. (Notice in contrast the correction scheme has smaller weights near the particle and larger weights away from the particle. ) Evidently, lower order schemes, for example trilinear, should perform better in a two-way coupled flows in the absence of a correction scheme.

We examine the Reynolds number dependence of the results for a settling particle in Figure 15. The magnitude of the steady state error decreases slightly with increasing Reynolds number. The unphysical intra-grid and inter-grid oscillations appear to reduce with increasing Reynolds number. The Reynolds number was changed by increasing the magnitude of gravity. This corresponds to a higher steady-state settling velocity and reduced intra-cell residence time. The reduced residance time means the particle samples a smaller distribtion of fluid velocity per unit time which may account for the reduction in oscillations.The general observation is that the correction scheme behaves well over the decade of Reynolds numbers examined.

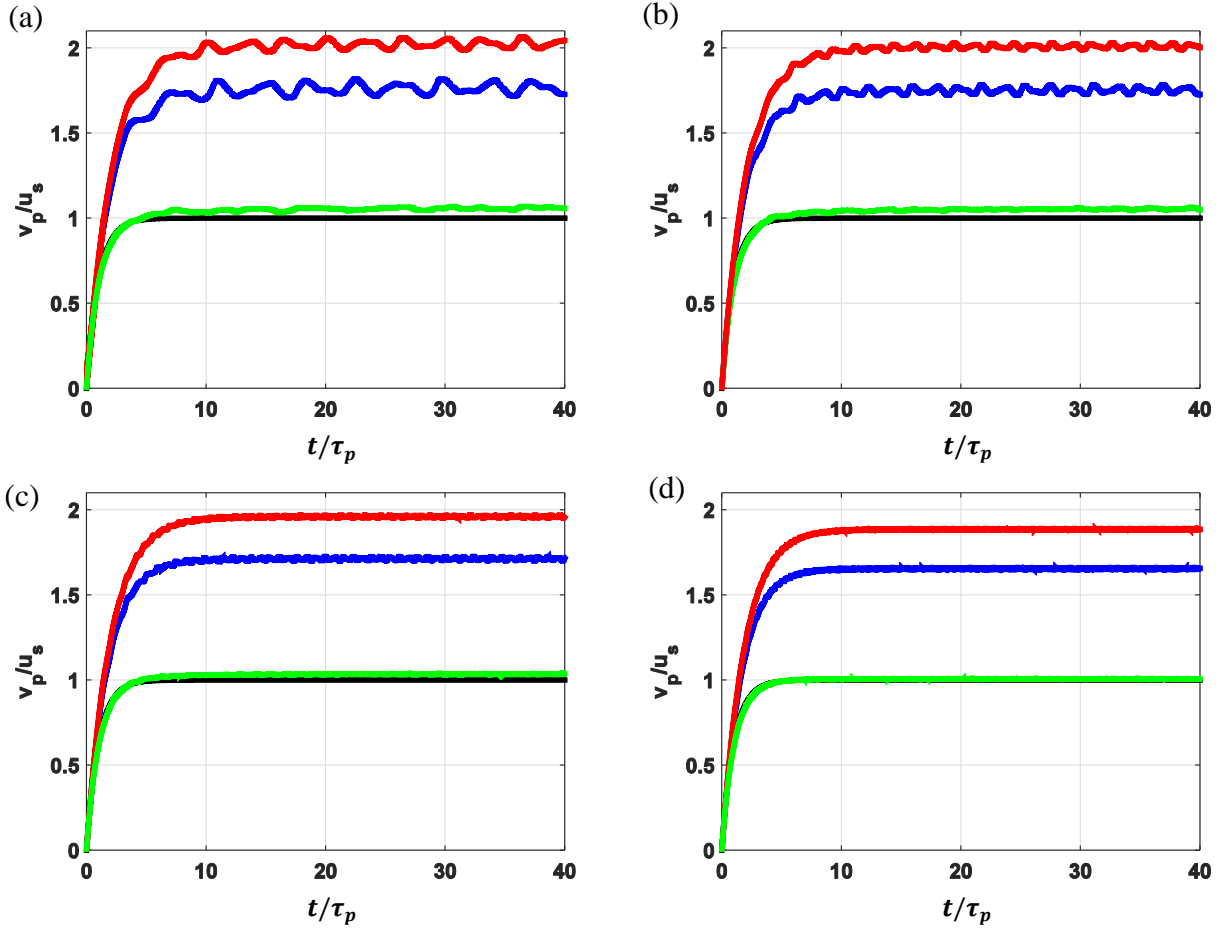

Figure 15: Particle transient settling velocity for different particle Reynolds numbers, red-4$^{th}$ order Lagrange, blue-2$^{nd}$ order trilinear interpolation, green-correction scheme, black-analytical solution, (a) $Re_p = 0.05$, (b) $Re_p = 0.10$, (c) $Re_p = 0.25$, (d) $Re_p = 0.50$, $\Lambda = 1$.

Another important parameter to examine is the Stokes number. In turbulent flows, the Stokes number plays an important role in the dictating the particle concentration field. With respect to estimating the undisturbed flow necessary to calculate the Stokes drag, there is a balance between the viscous relaxation time of the fluid due to the particle drag force and relaxation of the particle velocity due to the fluid force. The relaxation rates are respectively $\sim \exp(-t_{visc}/t)$ [1] and $\sim \exp(-t/t_p)$ so that for comparable ratio of particle to fluid time scales, e.g. $St = O(1)$, the time required for the disturbance field to develop in the fluid is significantly longer than the time it takes for a particle to assume a velocity comparable to the undisturbed fluid velocity. In other words, whereas the particle velocity analytically assumes 99% of it steady state velocity after ~5 particle relaxation times, the fluid disturbance field does develop to the same extant until ~100 viscous time scales. Therefore, we may expect that the correction scheme, which has $C$ coefficients tuned for steady state ($St \rightarrow \infty$) may not perform as well at low Stokes numbers. Nevertheless the proposed correction is shown to perform robustly across a wide range of Stokes numbers as shown in Figure 16.

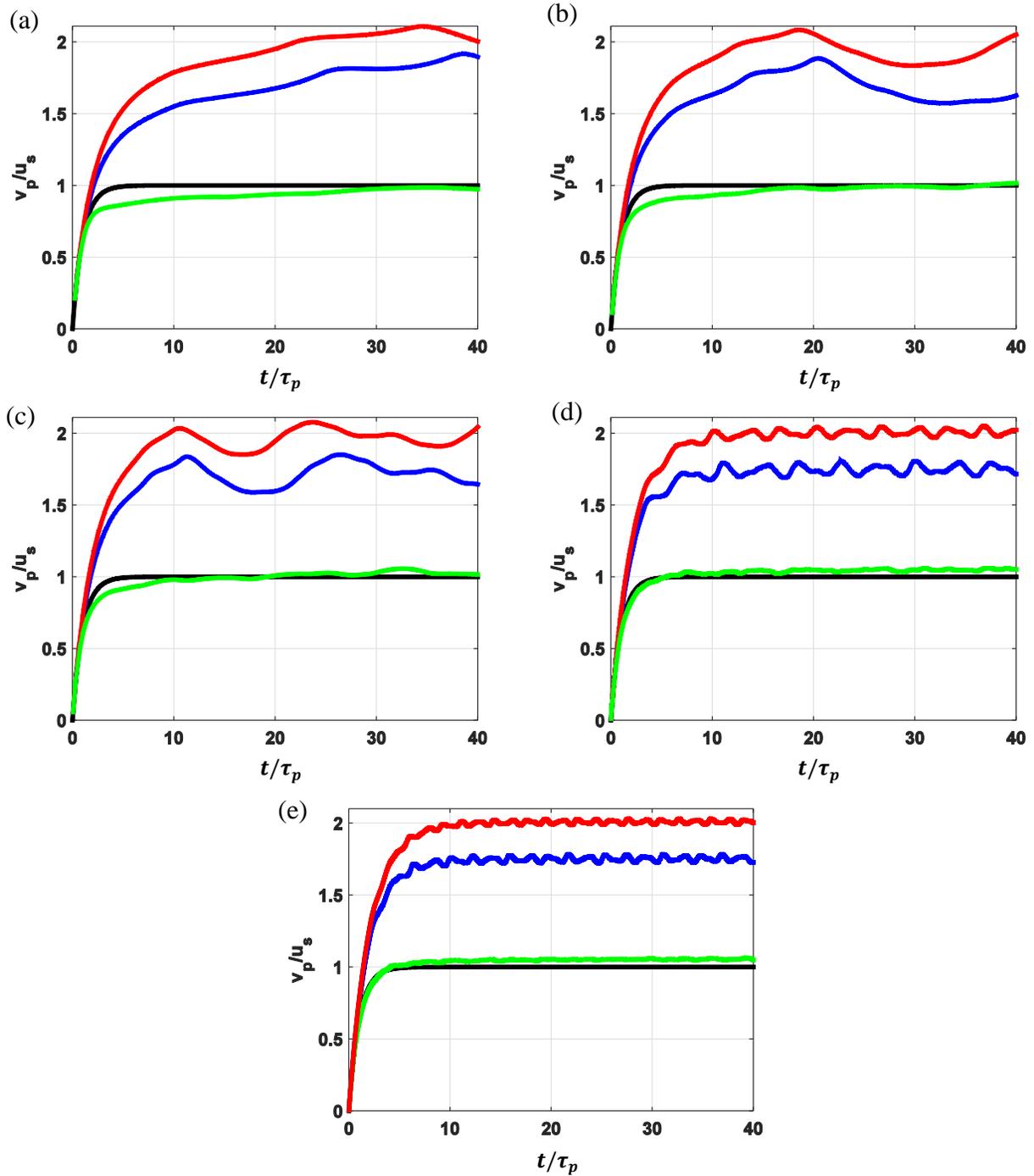

Figure 16: Transient particle settling velocity for different Stokes numbers (a) $St = 0.25$, (b) $St = 0.50$, (c) $St = 1.00$, (d) $St = 5.00$, (e) $St = 10.00$, $\Lambda = 1$, red-$4^{th}$ order Lagrange, blue-$2^{nd}$ order trilinear interpolation, green-correction scheme, black-analytical solution.

Transient settling velocity histories for different Stokes numbers are shown in Figure 16. For $St = 5, 10$, Figure 16 (d) and (e), the particle settling velocity is in excellent agreement with the analytical solution. For smaller Stokes numbers, there is some undershoot in the settling velocity predicted by the correction scheme, especially for $St = 0.25$. This is because the correction scheme is tuned for the steady problem of flow passed a sphere and is attempting to compensate for the amount of drag error that would occur in steady state. While the correction scheme has some unsteadiness built into it, that is, the Laplacian term grows as the disturbance flow develops, it seems for low Stokes numbers the correction scheme overpredicts the disturbance force error at early times. Hence, the drag force is overestimated at early times and the settling velocity is underestimated. Note that this is not a serious drawback since most physical problems consider fluid statistics averaged over several eddy turnover times which will be much longer than the viscous time scale. Note, even at low Stokes number, the correction particles approach the steady state analytical solution. In addition, the undershoots at early times are still considerably smaller than the overshoots that exist using trilinear and Lagrange-4 interpolation schemes without the proposed correction. The results presented in Figure 16 were for the maximum particle size and it is likely that errors at early times would be reduced for smaller particles. Based on these results, we may suggest that if the user of the correction scheme is interested in long time averaged statistics, then the proposed scheme should be appropriate for all of the size, Reynolds numbers, and Stokes numbers examined. If the user is interested in calculating short time dispersion statistics, they should proceed with caution for larger particles at small Stokes number.

Finally, we summarize the results of these numerical experiments of the settling particle in Tables 4-9. Here we present averaged settling and rms drift velocity statistics for each of the parameters we examined. Time averaging was performed over approximately thirty relaxation times from $t = 10\tau_p \rightarrow 40\tau_p$. While analytically there is no drift component in the particle trajectory (component perpendicular to gravity), there does end up being a drift component owing to interpolation errors as the particle moves in an arbitrary direction relative to the grid. In all of the previous figures, the settling velocity $u^{//} = \vec{v}_p \cdot \hat{g}$ is the component of the velocity parallel to the gravity vector and the drift velocity $u^+ = |\vec{v}_p - u^{//}\hat{g}|$ is the magnitude of the component perpendicular to the gravity vector. In all cases, the particle settling velocity for the corrected scheme exhibits at least one order of magnitude less error than the steady state settling velocity for either the trilinear or Lagrange-4 schemes. The maximum steady state error for the correction scheme is ~5.5% for $Re_p = 0.05, St = 10, \Lambda = 1.0$, which is signifcantly lower than errors for trilinear and Lagrange-4, ~76% and ~102% respectively. For all cases, trilinear interpolation was used to interpolate the C-field. Some tests using higher order interpolation (Lagrange-4) of the C field exhibited a little smaller error in drift velocity, but the variance in steady state settling velocity was not significantly different. While steady state settling velocity errors are small for all parameters tested, $\leq O(5\%)$, the non-zero error is likely due to small but finite asymmetry in the disturbance velocity field associated with a moving particle compared with a stationary particle.

Tables 7-9 show time averaged rms drift velocity normalized by $u_s = g\tau_p(1-\rho_f/\rho_p)$. For $St = 10$, for all particle sizes and Reynolds numbers, the particle drift velocity is small for all schemes, $O(1\%)$, although the correction scheme has lower error for most cases. The Lagrange-4 scheme has a little bit smaller error in drift velocity compared with the trilinear scheme. For Stokes numbers of unity and less, the drift velocity error becomes appreciable for the trilinear and Lagrange-4 schemes. For $St = 0.25$, Lagrange-4 exhibits ~9.2% time averaged rms error in drift velocity while trilinear exhibits almost 11% normalized drift velocity error. Note that for all parameters examined in this study, the normalized drift velocity error for the proposed correction scheme is less than 1%.

| Λ | 1.0 | 0.5 | 0.25 | 0.1 | 0.05 | 0.01 |
|---|---|---|---|---|---|---|
| Lagrange-4 | 101 | 50 | 24 | 8.5 | 3.5 | 0.26 |
| Lagrange-2 | 75 | 37 | 18 | 6.2 | 2.5 | 0.18 |
| Corrected | 5.1 | 2.4 | 1.0 | 0.35 | 0.16 | 0.016 |

Table 4: Percent Error in settling velocity, size study.

| $Re_p$ | 0.05 | 0.10 | 0.25 | 0.50 |
|---|---|---|---|---|
| Lagrange-4 | 102 | 101 | 96 | 89 |
| Lagrange-2 | 76 | 75 | 71 | 65 |
| Corrected | 5.5 | 5.1 | 3.4 | 0.65 |

Table 5: Percent Error in settling velocity, Reynolds study.

| St | 0.25 | 0.50 | 1.00 | 5.00 | 10.0 |
|---|---|---|---|---|---|
| Lagrange-4 | 99 | 95 | 97 | 100 | 101 |
| Lagrange-2 | 75 | 69 | 72 | 75 | 75 |
| Corrected | -4.7 | -1.7 | 1.4 | 4.5 | 5.1 |

Table 6: Percent Error in settling velocity, Stokes study.

| Λ | 1.0 | 0.5 | 0.25 | 0.1 | 0.05 | 0.01 |
|---|---|---|---|---|---|---|
| Lagrange-4 | 0.71 | 0.32 | 0.11 | 0.022 | 0.012 | 0.0050 |
| Lagrange-2 | 1.0 | 0.42 | 0.14 | 0.025 | 0.0058 | 0.0027 |
| Corrected | 0.22 | 0.063 | 0.020 | 0.010 | 0.011 | 0.0018 |

Table 7: Percent Error in rms drift velocity normalized by settling velocity, size study.

| $Re_p$ | 0.05 | 0.10 | 0.25 | 0.50 |
|---|---|---|---|---|
| Lagrange-4 | 1.3 | 0.71 | 0.32 | 0.20 |
| Lagrange-2 | 1.9 | 1.0 | 0.45 | 0.26 |
| Corrected | 0.37 | 0.22 | 0.13 | 0.093 |

Table 8: Percent Error in rms drift velocity normalized by settling velocity, Reynolds study.

| St | 0.25 | 0.50 | 1.00 | 5.00 | 10.0 |
|---|---|---|---|---|---|
| Lagrange-4 | 9.2 | 5.8 | 4.7 | 1.4 | 0.71 |
| Lagrange-2 | 11 | 7.3 | 6.1 | 2.0 | 1.0 |
| Corrected | 0.86 | 0.89 | 0.61 | 0.39 | 0.22 |

Table 9: Percent Error in rms drift velocity normalized by settling velocity, Stokes study.

## 5.1 Energetics

Finally, it is worth mentioning what are the consequences of using the proposed scheme for computing the particle drag force on the energetics of the particles and fluid separately as well as for the system as a whole. At the outset it is clear that the improved prediction of particle kinematics by the proposed correction scheme will directly translate to improved prediction of particle and fluid energetics. For example, improved prediction of the particle settling velocity implies an improved prediction of the rate at which potential energy is being consumed.

To develop more insight into energy exchange processes between the particle and the fluid, we next present the fluid and particle energy equations. Our derivations closely follow the work of Sundaram and Collins [28]. The particle kinetic energy equation is readily obtained from (5):

$$(11) \quad \frac{dk_p}{dt} = v_i m_p (\tilde{u}_i - v_i)/\tau_p + m_p(1 - \rho_f/\rho_p) g_i v_i$$

Where $k_p = \frac{1}{2} m_p v_i v_i$ is the particle kinetic energy. The fluid kinetic energy equation is obtained from the Navier-Stokes equation (2) viz.

$$(12) \quad u_i \times \left[ \frac{\partial}{\partial t} \rho_f u_i + \frac{\partial}{\partial x_j} \rho_f u_j u_i = -\frac{\partial p}{\partial x_i} + \mu \frac{\partial^2 u_i}{\partial x_j \partial x_j} - \frac{1}{V} F_{d,i} P\{\delta(x_i - x_i')\} \right]$$

$$(13) \quad \frac{\partial k}{\partial t} + \frac{\partial}{\partial x_j} u_j k = -u_i \frac{\partial p}{\partial x_i} + \mu u_i \frac{\partial^2 u_i}{\partial x_j \partial x_j} - \frac{1}{V} u_i F_{d,i} P\{\delta(x_i - x_i')\}$$

Where $k_f = \left(\frac{1}{2}\right) \rho_f u_i u_i$. Integrating over the volume, the convective and pressure work terms disappear since they can be written in divergence form yielding[3]:

$$(14) \quad \frac{dk_f}{dt} = W_v - \epsilon_{pp}$$

Where:

$$(15) \quad W_v = \int \mu u_i \frac{\partial^2 u_i}{\partial x_j \partial x_j} dV = \int \left[ \mu \nabla^2 k - \epsilon + \mu \frac{\partial u_i}{\partial x_j} \frac{\partial u_j}{\partial x_i} \right] dV$$

---

[3] This has been verified for our staggered scheme but in general this step is contingent upon the discrete operators' satisfaction of the rules of calculus.

$$\text{(16)} \quad \epsilon_{pp} = \int \frac{1}{V} u_i F_{d,i}\, P\{\delta(x_i - x_i')\}dV = u_i m_p(\tilde{u}_i - v_i)/\tau_p$$

Here $\epsilon = 2\mu S_{ij}S_{ij}$ is the true fluid dissipation rate per unit volume, where $S_{ij} = \left(\frac{1}{2}\right)\left(\frac{\partial u_i}{\partial x_j} + \frac{\partial u_j}{\partial x_i}\right)$ is the strain-rate tensor.

Summing expressions (11) and (14), the evolution of the total kinetic energy of the system is described by (17):

$$\text{(17)} \quad \underbrace{\frac{dk_p}{dt} + \frac{dk_f}{dt}}_{\text{Change in KE}} = \underbrace{m_p \frac{(v_i - u_i)(\tilde{u}_i - v_i)}{\tau_p} + W_v}_{\text{Heat Generation}} + \underbrace{m_p\left(1 - \frac{\rho_f}{\rho_p}\right)g_i v_i}_{\text{Work In}}$$

The first term on the right hand side of (17) is owing to the incorporation of a point-particle model and would be absent in a particle resolved simulation. In the case of point-particle simulations, this term represents a portion of the heat generation due to unresolved particle boundary layers on the particles. Note that in the limit that $\Lambda \to 0$, where $\tilde{u}_i \approx u_i$, (17) reduces to the kinetic energy equation derived in [28]. The only difference lies in the point-particle source term, the first term on the right hand side of (17). We note that based on the analytical solution to the Stokes equation, the rate of heat generation must be $m_p(\tilde{u} - v_p)^2/\tau_p$, however this is not explicitly seen in the expression above. To clarify this connection, we re-write the heat generation term in (17) in the following form:

$$\text{(18)} \quad m_p \frac{(v_i - u_i)(\tilde{u}_i - v_i)}{\tau_p} + W_v = -\frac{m_p(\tilde{u}_i - v_i)^2}{\tau_p} + \frac{m_p(\tilde{u}_i - u_i)(\tilde{u}_i - v_i)}{\tau_p} + W_v$$

The first term on the right hand side of (18) represents the analytical heat generation. Using (14) and (16) and assuming the scheme is predicting $\tilde{u}_i = 0$, one can verify that the second and third term on the right hand side of (18) will cancel in steady-state. The above exercise clarifies how each of the respective terms should be interpreted. For example, in the case of refinement to very small mesh for a fixed particle size, the disturbed fluid velocity may be singularly large leading to artificial apparent viscous dissipation. This is an inevitable consequence of the use of a point-force model. However, the second term on the right hand side of (18) grows in the same fashion with the opposite sign and will cancel this unphysical effect.

Our discussion has focused mostly on the consequences of implementing a verifiable Stokes drag scheme within a point-particle code with respect to kinematics and energetics for low particle Reynolds numbers. Provided the assumptions of Stokes drag are met, it is expected that the demonstrated improvements can be translated to more complex scenarios such as turbulent flows laden with particles. While we defer investigation of such settings to a future publication, we briefly note that previous numerical studies by Segura [12] and Yamamoto [34] in channel flow found that turbulence attenuation by particles was significantly under-predicted compared with experiments under similar conditions performed by Paris [2] and Kulick et al. [13] respectively. Our results may explain some of these discrepancies by noting that under-prediction of the drag force by an uncorrected point-particle model will naturally lead to under-prediction of the energy exchange between particles and fluid. This under-prediction can be significant, especially in settings with near-wall mesh refinement where most turbulence dissipation takes place. In these regions, $\Lambda$ will inevitably be large.

A final word concerns the utility of point-particle methods. It should be emphasized that the mere use of a point-force model implies that the fluid structures in the neighborhood of particles are necessarily artificial. The premise of a verifiable point-force model is to ensure that the flow structures away from the particle "see" the correct momentum input. Under conditions where the Stokes drag and gravity force are not the only leading order interactions between the particle and fluid, other terms in the Maxey-Riley equation or other equation of motion will have to be used. These equations of motion have other terms (e.g. added mass, history) which also depend on undisturbed fluid information in the same fashion as the Stokes drag. Ideas similar to those presented here are likely to lead to improved predictions in these regimes as well. Nevertheless, careful verification benchmarks must be conducted to establish the effectiveness of similar correction schemes for more complicated equations of motion.

## 6. Summary

In this work, we have demonstrated the error that is incurred in the calculation of Stokes drag in particle laden flows when the incorrect velocity, namely the disturbed fluid velocity is used in place of the correct, undisturbed fluid velocity. For uniform flow passed a stationary point-particle, the error is demonstrated to be the same order as the particle size compared with the grid spacing. Projecting the particle force over more grid points can reduce but not eliminate the drag force error. Only a projection kernel of infinite support could accomplish that.

A correction scheme is proposed to estimate the undisturbed fluid velocity from the neighboring disturbed fluid velocity information. The correction scheme may be interpreted as a non-convex interpolation scheme with negative weights near the particle and positive weights away from the particle. This allows the scheme to estimate a velocity which is greater than any of the velocity data used in the scheme (for flow passed a stationary particle) which is in the direction that the curvature vector points. The scheme was extended to an arbitrary location of the particle embedded in an Eulerian grid. For a given $\Lambda$, the scheme is well represented by six $C$ constants which are independent of time and identical for all grid cells. For any particle location in the grid, the $C$ field may be interpolated to the particle location. Whereas simple interpolation of the fluid velocity field is not directly appropriate for two-way coupling since the velocity field is contaminated owing to the disturbance field that a particle generates for itself, the $C$ field is not contaminated during the simulation. For particle sizes that do not correspond to the six tested in this work, it is straightforward to evaluate the spline equations for the particle sizes chosen for a particular physical study. These spline interpolated coefficients will become the six coefficients that are then interpolated at each time step during simulation runtime based on a particle's position in a grid cell.

Having developed the correction scheme for flow passed a stationary particle, the scheme is tested for a particle settling in an arbitrary direction relative to the grid at low Reynolds number. The correction scheme to calculate the undisturbed fluid velocity is compared with standard interpolation schemes for calculating the disturbed fluid velocity at the particle location. We compared these schemes for different particle sizes, Reynolds numbers, and Stokes numbers. For all cases, the correction scheme was superior compared with trilinear and Lagrange-4 schemes. Time-averaged steady state settling velocity error was at least one order of magnitude smaller for the correction scheme compared with the other schemes. In addition, the correction scheme was shown to significantly suppress the unphysical inter-grid and intra-grid oscillations compared with the interpolation schemes. The calculated drift velocity was small for all schemes except at small Stokes number where the interpolation schemes exhibited drift velocity error of $O(10\%)$. Another important finding was that the trilinear scheme had lower error in settling velocity compared with the Lagrange-4 scheme. This is because the Lagrange-4 scheme provides a better estimate of the disturbed flow velocity at the particle, which is a worse estimate of the undisturbed fluid velocity. Further, the Lagrange-4 scheme has larger interpolation weights near the particle so that projection of the interpolated force tends to exacerbate the error. Therefore, an important finding is that higher order interpolation schemes will tend to perform worse than lower order interpolation schemes.

This work moves the point-particle method towards being a verifiable scheme. Accurate calculation of the Stokes drag means Lagrangian statistics including particle distribution and dispersion will be more accurate. We have demonstrated that interpolation-projection with un-corrected schemes will under-predict the momentum exchange between particles and fluid in simulations of turbulent flows laden by particles. This is because un-corrected schemes use the difference between the disturbed and particle velocity as the slip velocity to calculate the Stokes drag force. This quantity will be less than if the correct slip velocity were used, namely the difference between the undisturbed velocity which is $O(u'_{rms})$ and the particle velocity. This is a significant observation as it may elucidate why point-particle simulations of turbulence disagree with experimental efforts with regard to turbulence modification by particles. It is also worth further investigation to reveal how different numerical methods for point-particles influence the prediction of Eulerian and Lagrangian statistics in two-way coupled turbulent flows.

The correction scheme presented here was developed strictly for *momentum coupling* in dilute particle-laden flows and is appropriate whenever the computational user can justify the Stokes drag should be the leading order contribution, (for instance, high density ratio, $d_p/\eta < 1$, $Re_p < 1$, etc.). The scheme was implemented on a uniform staggered grid using a 2$^{nd}$ order discretization. The method may be suitable for higher order discretization since the $C$ field was calculated via a thorough grid refinement study. Extensions to non-uniform grids as would be found in channel flows is straightforward, but will require some modification since the assumed symmetry in the C field is partially broken. (For non-uniform grids, eight coefficients instead of six would be necessary.)

**Acknowledgement**
This work is based on results presented at the 2014 APS DFD. This work was supported by the United States Department of Energy under the Predictive Science Academic Alliance Program 2 (PSAAP2) at Stanford University. Jeremy Horwitz was supported by the National Science Foundation Graduate Research Fellowship under Grant No. DGE-114747. Any opinion, findings, and conclusions or recommendations expressed in this material are those of the authors(s) and do not necessarily reflect the views of the National Science Foundation.

**Appendix: Spline Interpolants for C-field**

The equation for the $i^{th}$ $C$ coefficient, $C_i$, $i: 1 \to 6$, in the $j^{th}$ $\Lambda$ interval, $j: 1 \to 5$ is given in equation (A.1), constructed by a cubic spline approach:

$$(A.1) \qquad C_i(\Lambda) = C_{ij4} + C_{ij3}(\Lambda - \Lambda_{oj}) + C_{ij2}(\Lambda - \Lambda_{oj})^2 + C_{ij1}(\Lambda - \Lambda_{oj})^3$$

Here, $\Lambda_{jo}$ is a virtual origin for the $j^{th}$ $\Lambda$ interval, viz.

$$(A.2) \qquad \Lambda_{jo} = [0.01, 0.05, 0.1, 0.25, 0.5]$$

The $j$ $\Lambda$ intervals are:

$$(A.3) \qquad \Delta\Lambda_j = \{[0.01, 0.05], \quad [0.05, 0.1], \quad [0.1, 0.25], \quad [0.25, 0.5], \quad [0.5, 1.0]\}$$

(A.1) represents a smooth interpolation of the $C$ field given in Table 2 and reproduced below in Table A.1. The $k^{th}$ coefficient, $C_{ijk}$ for the $i^{th}$ scalar value $C_i$, in the $j^{th}$ $\Delta\Lambda$ interval are given in Tables A.2-A.7 below.

| $\Lambda$ | $C_1$ | $C_2$ | $C_3$ | $C_4$ | $C_5$ | $C_6$ |
|---|---|---|---|---|---|---|
| 1.0 | 0.248 | 0.312 | 0.340 | 0.440 | 0.681 | 0.518 |
| 0.5 | 0.246 | 0.309 | 0.337 | 0.435 | 0.671 | 0.512 |
| 0.25 | 0.242 | 0.303 | 0.329 | 0.422 | 0.644 | 0.493 |
| 0.1 | 0.231 | 0.285 | 0.306 | 0.389 | 0.579 | 0.447 |
| 0.05 | 0.218 | 0.267 | 0.282 | 0.356 | 0.522 | 0.403 |
| 0.01 | 0.190 | 0.236 | 0.239 | 0.311 | 0.460 | 0.335 |

Table A.1: $C$-field coefficients for different values of $\Lambda$

|       | $k=1$              | $k=2$               | $k=3$              | $k=4$              |
|-------|--------------------|---------------------|--------------------|--------------------|
| $j=1$ | 29.7380852550663   | -8.75483997204750   | 1.00261266247379   | 0.190000000000000  |
| $j=2$ | 29.7380852550664   | -5.18626974143956   | 0.444968273934312  | 0.218000000000000  |
| $j=3$ | 1.45733053808526   | -0.725556953179597  | 0.149376939203354  | 0.231000000000000  |
| $j=4$ | 0.0537582110412291 | -0.0697582110412291 | 0.0300796645702305 | 0.242000000000000  |
| $j=5$ | 0.0537582110412291 | -0.0294395527603073 | 0.00528022361984637| 0.246000000000000  |

Table A.2: $C_{1jk}$

|       | $k=1$              | $k=2$               | $k=3$              | $k=4$              |
|-------|--------------------|---------------------|--------------------|--------------------|
| $j=1$ | 24.7650593990219   | -7.83056883298398   | 1.04859865828092   | 0.236000000000000  |
| $j=2$ | 24.7650593990218   | -4.85876170510134   | 0.541025436757512  | 0.267000000000000  |
| $j=3$ | 2.25392033542974   | -1.14400279524807   | 0.240887211740041  | 0.285000000000000  |
| $j=4$ | 0.105738644304684  | -0.129738644304684  | 0.0498259958071283 | 0.303000000000000  |
| $j=5$ | 0.105738644304684  | -0.0504346610761710 | 0.00478266946191452| 0.309000000000000  |

Table A.3: $C_{2jk}$

|       | $k=1$              | $k=2$               | $k=3$              | $k=4$              |
|-------|--------------------|---------------------|--------------------|--------------------|
| $j=1$ | 36.3508036338224   | -11.3367155835080   | 1.47030733752620   | 0.239000000000000  |
| $j=2$ | 36.3508036338226   | -6.97461914744934   | 0.737853948287911  | 0.282000000000000  |
| $j=3$ | 3.04933612858142   | -1.52199860237596   | 0.313023060796646  | 0.306000000000000  |
| $j=4$ | 0.115130677847659  | -0.149797344514326  | 0.0622536687631028 | 0.329000000000000  |
| $j=5$ | 0.115130677847659  | -0.0634493361285815 | 0.00894199860237598| 0.337000000000000  |

Table A.4: $C_{3jk}$

|       | $k=1$              | $k=2$               | $k=3$              | $k=4$              |
|-------|--------------------|---------------------|--------------------|--------------------|
| $j=1$ | 18.4394129979040   | -7.56379035639415   | 1.39804855345912   | 0.311000000000000  |
| $j=2$ | 18.4394129979034   | -5.35106079664568   | 0.881454507337526  | 0.356000000000000  |
| $j=3$ | 5.47236897274636   | -2.58514884696018   | 0.484644025157233  | 0.389000000000000  |
| $j=4$ | 0.0665828092243163 | -0.122582809224316  | 0.0784842767295594 | 0.422000000000000  |
| $j=5$ | 0.0665828092243162 | -0.0726457023060791 | 0.0296771488469605 | 0.435000000000000  |

Table A.5: $C_{4jk}$

|       | $k=1$               | $k=2$               | $k=3$              | $k=4$              |
|-------|---------------------|---------------------|--------------------|--------------------|
| $j=1$ | -0.442208245981379  | -4.49806848357797   | 1.73063027253669   | 0.460000000000000  |
| $j=2$ | -0.442208245981526  | -4.55113347309576   | 1.36866219426974   | 0.522000000000000  |
| $j=3$ | 9.58758909853245    | -4.61746470999300   | 0.910232285115303  | 0.579000000000000  |
| $j=4$ | 0.185716282320059   | -0.303049615653393  | 0.172155136268345  | 0.644000000000000  |
| $j=5$ | 0.185716282320059   | -0.163762403913348  | 0.0554521313766593 | 0.671000000000000  |

Table A.6: $C_{5jk}$

|       | $k=1$              | $k=2$               | $k=3$              | $k=4$              |
|-------|--------------------|---------------------|--------------------|--------------------|
| $j=1$ | 43.5868623340319   | -14.7774032145352   | 2.22135714884696   | 0.335000000000000  |
| $j=2$ | 43.5868623340325   | -9.54697973445148   | 1.24838183088749   | 0.403000000000000  |
| $j=3$ | 6.10772886093639   | -3.00895038434660   | 0.620585324947588  | 0.447000000000000  |
| $j=4$ | 0.175139063591895  | -0.260472396925228  | 0.130171907756814  | 0.493000000000000  |
| $j=5$ | 0.175139063591894  | -0.129118099231307  | 0.0327742837176799 | 0.512000000000000  |

Table A.7: $C_{6jk}$